\documentclass[aps,pra,showpacs,amsfonts,amsmath,amssymb,reprint,twocolumn,superscriptaddress,floatfix]{revtex4-1}

\usepackage{graphicx}% Include figure files
\usepackage[utf8]{inputenc}
\usepackage[T1]{fontenc}
\usepackage{color}
\usepackage[colorlinks,bookmarks=false,citecolor=blue,linkcolor=red,urlcolor=black]{hyperref}
\usepackage{ulem}
\newcommand{\be}{\begin{equation}}
\newcommand{\ee}{\end{equation}}
\newcommand{\bea}{\begin{eqnarray}}
\newcommand{\eea}{\end{eqnarray}}
\newcommand{\da}{\dagger}

\newcommand{\av}[1]{\left\langle #1\right\rangle}
\newcommand{\Var}[0]{\mathrm{Var}}

\newcommand{\Tr}{\textrm{Tr}}

\newcommand{\rme}{\mathrm{e}}
\newcommand{\rmi}{\mathrm{i}}
\newcommand{\rmd}{\mathrm{d}}

\newcommand{\avnsq}{\langle n_i^2 \rangle}
\newcommand{\avn}{\langle n_i \rangle}

\newcommand{\cd}{c^{\dagger}}
\newcommand{\dd}{d^{\dagger}}
\newcommand{\Ad}{A^{\dagger}}
\newcommand{\Bd}{B^{\dagger}}
\newcommand{\bessel}{{\cal I}_0}
\newcommand{\nqp}{n^{\mathrm{(qp)}}}
\newcommand{\nqh}{n^{\mathrm{(qh)}}}
\newcommand{\vqp}{v^{\mathrm{(qp)}}}
\newcommand{\vqh}{v^{\mathrm{(qh)}}}
\newcommand{\jqp}{j^{\mathrm{(qp)}}}
\newcommand{\jqh}{j^{\mathrm{(qh)}}}
\newcommand{\jE}{j^E}
\newcommand{\jN}{j^n}
\newcommand{\jn}{j^n}

\begin{document}
\title{Dynamics of heat and mass transport in a quantum insulator}

\author{Mateusz \L{}\k{a}cki}
\affiliation{Instytut Fizyki imienia Mariana Smoluchowskiego, Uniwersytet Jagiello\'nski, ulica \L{}ojasiewicza 11
30-348  Krak\'ow, Poland}
\affiliation{Laboratoire Kastler Brossel, UPMC-Sorbonne Universit\'es, CNRS, ENS-PSL Research University, Coll\`ege de France,
4 place Jussieu 75005 Paris, France}
\author{Dominique Delande}
\affiliation{Laboratoire Kastler Brossel, UPMC-Sorbonne Universit\'es, CNRS, ENS-PSL Research University, Coll\`ege de France,
4 place Jussieu 75005 Paris, France}
\author{Jakub Zakrzewski}
\affiliation{Instytut Fizyki imienia Mariana Smoluchowskiego, Uniwersytet Jagiello\'nski, ulica \L{}ojasiewicza 11
30-348  Krak\'ow, Poland}
\affiliation{Mark Kac Complex Systems Research Center, Jagiellonian University, ulica \L{}ojasiewicza 11
30-348 Krak\'ow, Poland}

\date{\today}


\begin{abstract}
The real time evolution of two pieces of quantum insulators, initially at different temperatures, is studied when they are glued together.
Specifically, each subsystem is taken  as a Bose-Hubbard model in a Mott insulator state. The process of temperature
equilibration via heat transfer is simulated in real time using the Minimally Entangled Typical Thermal States algorithm. The analytic
theory based on quasiparticles transport is also given.
\end{abstract}
\maketitle

%%%%%%%%%%%%%%
%INTRODUCTION
%%%%%%%%%%%%%%
\section{Introduction}

One of the open problems of statistical mechanics is the description of isolated quantum many-body systems far from equilibrium.  
Simulations of real time dynamics are quite difficult and requiring large computer resources even for the simplest nontrivial 
one-dimensional systems with short range interactions, although a significant progress has been obtained in {the} last ten or so years when
modern versions of Density Matrix Renormalization Group (DMRG) \cite{White92} approaches using  Matrix Product States (MPS) have been developed
\cite{Vidal03,Vidal04} - for recent reviews see \cite{Schol05,Schol11}. It has been understood that the growth of entanglement 
in the system is a main obstacle in studying long-time dynamics even for initial states being close to the ground state, for which the 
entanglement is typically quite low \cite{Schol11}. This growth may be {even} more restrictive for abrupt quenches and study of excited states \cite{Lacki12}.

By comparison, the study of finite temperature dynamics is much less understood. Some results for large systems were obtained assuming adiabatic evolution (with entropy conservation \cite{Ho07,Pollet08}). One possible ``quasi-exact'' approach involves MPS of a system enlarged by an auxiliary environment. By tracing out the auxiliary degrees of freedom, one obtains the density matrix
of the original system \cite{Verstraete04,Zwolak04}. This approach has been applied for dynamics of  thermal systems \cite{Feiguin05,Barthel09,Feiguin11}.
Interestingly, there is an important freedom in the approach, namely that the same density matrix of the system may be obtained for different dynamics of the environment.
One may use this freedom in the wavefunction describing the system and the environment, in order to minimize the growth of system-environment entanglement during the temporal evolution.
That in turn allows for reaching longer times of evolution~\cite{Karrasch12,Karrasch13}. This approach has been used to calculate particle currents and Drude weight as well as spin correlations in XXZ spin chains. 
 
Another promising approach is based on the combination of time-dependent DMRG in the Heisenberg picture -- that allows to obtain the real time evolution of the operator~\cite{Prosen07} -- with the density matrix obtained using imaginary time propagation. This allows for an evaluation of expectation values on a grid of temperature/time points simultaneously \cite{Pizorn14}.
 
 We shall use in the present paper yet another approach called Minimally Entangled Typical Thermal State (METTS) approach 
 \cite{White09,Stoudenmire10,Metts14}. It has been argued that it can simulate finite temperature quantum systems with a computational cost comparable to ground state DMRG \cite{Stoudenmire10}.  As its application to real time dynamics requires propagation of excited states for which the entanglement growth may be a serious obstacle \cite{Prosen09}, this method may be costly to implement. 
 A very recent study \cite{Barthel14} has shown that METTS may be quite efficient for gaped systems at low temperatures. We shall use it for studying the 
dynamics of gaped Mott insulators far from equilibrium. 
 
 We will mainly consider a system composed of two one-dimensional Mott insulators at different temperatures that are glued together at a given instant of time.
 Such a system can, in principle, be realized in the laboratory. A single insulator is routinely observed by placing ultra-cold 
atoms in a one-dimensional deep optical lattice with the transverse directions being  frozen by a tight laser confinement \cite{Esslinger04}. 
The system can be split by an additional laser potential into two separate parts. One may imagine heating both parts differently and then bringing them into contact by a rapid switch-off of the separating laser.
 
 For such a model system, we concentrate on the heat transfer assuming little direct particle current. In Section II, we discuss briefly our METTS implementation for the Bose-Hubbard Hamiltonian and show exemplary results. Section III brings a simple analytical theory for the observables assuming that the heat transport is dominated by quasiparticle motions,
 described (in the low tunneling limit) by the Bogoliubov approach. Both approaches are compared in Section IV where we also discuss an alternative approach {in which} the two subsystems
are smoothly glued. We conclude in Section V.

\section{The system studied using METTS}

\subsection{METTS algorithm}

The METTS algorithm as proposed by White \cite{White09} simulates a thermal canonical ensemble with temperature $T=1/\beta$ (from now on we shall assume the convenient units with the Boltzmann constant  $k_B=1$, we assume also $\hbar=1$). 
Since the method is described in detail elsewhere \cite{Stoudenmire10},
we provide essentials 
 only. The METTS approach works by alternately applying the imaginary time evolution operator $\exp(-\beta H/2)$ and a projection measurement onto a given basis set. 
This defines a random walk which samples the Hilbert space and creates a properly weighted ensemble $\mathcal{M}$ which enables to estimate thermal averages of operators in the canonical ensemble: $ \langle O \rangle = \Tr [\exp(-\beta H) O]$ as $ \langle O \rangle = \sum\limits_{\psi \in \mathcal{M}} \langle \psi | O | \psi \rangle.$
The METTS ensemble allows also  for real time evolution of the ensemble $ \langle O(t) \rangle=\sum\limits_{\psi \in \mathcal{M}} \langle \psi | O(t) | \psi \rangle=\sum\limits_{\psi \in \mathcal{M}} \langle \psi(t) | O | \psi(t) \rangle.$ Both the imaginary time and the real time evolutions are performed efficiently with the time-evolving block decimation (TEBD) algorithm \cite{Vidal03}, essentially equivalent to a time-dependent DMRG approach.

The Hamiltonian of the one-dimensional Bose-Hubbard (BH) model reads 
\be
H = \sum_i H_i= \sum\limits_{i=1}^L \left \{ -J_i\left [ a_i a_{i+1}^\dagger + h.c \right ]+ \frac{U}{2}n_i(n_i-1) \right \} 
\label{eqn:BH}
\ee
with $a_i$ ($a_i^\dagger$) being the standard boson annihilation (creation) operator at site $i=1,...,L$, $n_i=a_i^\dagger a_i$, $J_i=J$ is the amplitude of hopping (tunneling)
between $i$ and $i+1$ site; $J_L=0$ for open boundary conditions (OBC) considered below 
while $U$ denotes the interaction strength. 
We consider the insulating regime $U\gg J$ (we shall assume typically that $J/U=0.05$ deep in the Mott regime).
In the METTS approach, the projection measurement can be performed on any basis set. It is here convenient to use a Fock basis
on each lattice site, meaning that we measure the (random) number of particle on each site, an easy task 
for a MPS state~new{\cite{Delande:BrightSoliton:NJP13}}. For low temperatures
considered in this work (we take a typical $\beta U$ of the order of 5), 
successive evolution steps between measurements are correlated. To
avoid that, we have started the METTS algorithm from the ground state neglecting the first 500 iterations of the METTS sampling procedure. For the final ensemble, we have taken every 200th METTS vector. Such a sampling is quite time consuming. The alternative would be to change the basis in which the measurements are performed. That, however, would significantly slow down
the time evolution using TEBD, because it would break the total particle number conservation (which would of course
be restored after statistical averaging)
\cite{Vidal03}.

\subsection{System preparation}

Our aim is to study the transport in a non-equilibrium system of two insulators at different temperatures brought into contact at a given time. Two
versions of the procedure will be considered:
\begin{itemize}
\item Two uniform systems with different temperatures are merged by building the tensor product of their density matrices.
\item Preparation of a temperature-inhomogeneous system with a smooth temperature gradient across the sample,
using the canonical thermal distribution of an auxiliary Hamiltonian.
\end{itemize}

\subsubsection{Simple tensor product approach}
\label{sec:tp}
To prepare the initial situation, we have considered two lattices of the same length $M=35$ sites that are linked together forming a longer lattice of length $L=2M.$ The Hamiltonian is  $H=H_L+H_R+V_{LR}$ where both $H_i$ terms  take the form (\ref{eqn:BH}) and
$V_{LR}$ is the hopping term connecting the two lattices.
Let us define a nonstandard but  useful convention that the two systems are linked ``symmetrically'' at $i=0$.
Thus the indices of the right hand side lattice of $M$ sites take half integer values $i=1/2,3/2,...,M/2-1/2$ while those of the left hand side lattice are
$i=-M/2+1/2,...-3/2, -1/2$. Then the coupling term between two subsystems reads $-J(a_{-1/2}a^\dagger_{1/2} +h.c.)$. We shall 
assume this ``symmetric'' notation from now on.

 The density matrix that describes the full system would be just the product of constituents' density matrices: $\rho=\rho_L \otimes \rho_R,$ each prepared separately under open boundary conditions at different temperatures, i.e.,  $\rho_i=\exp(-\beta_i H_i$) with $i=L,R$.
 We compute  two appropriate METTS $\mathcal{M}_L$ and $\mathcal{M}_R$ that represent  the density matrices $\rho_L$ and $\rho_R.$ The METTS ensemble that represents the density matrix $\rho_L\otimes \rho_R$ is just $\mathcal{M}_L \otimes \mathcal{M}_R := \{ \psi_L \otimes \psi_R | \psi_i \in \mathcal{M}_i, i=L,R\}.$
This is not efficient computationally. For example when estimating $\langle O_i \rangle$ for $i\in\{L, R\}$ the actual average is as follows (in this case for $L$):
$\langle \psi_L | O_i | \psi_L \rangle\langle \psi_R | \psi_R \rangle = \langle \psi_L |O_i | \psi_L \rangle.$ The sum over METTS vectors from $\mathcal{M}_L \otimes \mathcal{M}_R$ will contain repetitions of each average. It will occur as many times as many vectors are in $\mathcal{M}_R.$ On the other hand, if the two METTS ensembles contain exactly the same number of vectors $V$, then the METTS ensemble $\mathcal{M}_L \otimes \mathcal{M}_R$ yields completely equivalent estimates as: $\mathcal{M}_L \oplus \mathcal{M}_R:=\{\psi_L^i\otimes\psi_R^i | i=1,\ldots,V \},$ where $\psi_{L/R}^i$ denotes the $i$-th METTS vector of $\mathcal{M}_{L/R}.$ 

In typical applications to the BH model, we have found that the size of $\mathcal{M}_L \otimes \mathcal{M}_R$ is several millions of METTS which is computationally prohibitive, while $\mathcal{M}_L \oplus \mathcal{M}_R$ contains only the same number of vectors as each of $\mathcal{M}_L$ and $\mathcal{M}_R.$
We have therefore used the second approach.

\begin{figure}
\includegraphics[width=8cm]{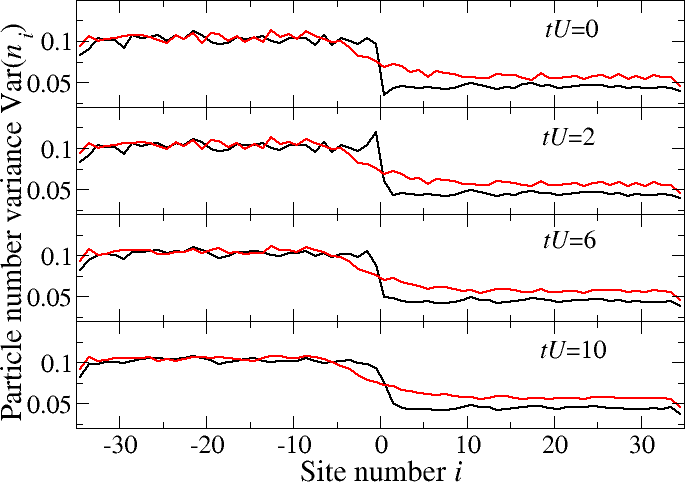}
\caption{Dynamical evolution of the variance of the particle number on each site. The initial state is chosen with an inhomogeneous temperature profile, in order to induce an heat current. The black curves are obtained for a tensor product initial state, where the left and right parts of the system are both at thermal equilibrium, but at different temperatures $\beta_L U=6,\beta_R U=8.$ 
The red curves correspond to "smoothly" glued initial conditions  as described in the main text. For $tU=2,$ the cusp in the black curve is a remnant of boundary effects occurring when tunneling between the two subsystems is abruptly turned on at $t=0.$ The short range fluctuations on the curves 
are of statistical origin, an unavoidable effect of the METTS approach. They vanish when the number of METTS states tends to infinity.}
\label{fig:fluct}
\end{figure}

The Hamiltonian governing the dynamics of the whole lattice couples the ``left'' and ``right'' parts of the system through the additional tunneling term, mentioned above,  $-Ja_{-1/2}a_{1/2}^\dagger + h.c.$. 
When using the tensor product of two density matrices as an initial state, the gluing region is subject to violent short-time transient phenomena 
as shown in Fig.~\ref{fig:fluct}: see for example the increased variance just on the left of the gluing point at $tU=2$. They are the remnants of the OBC for the two constituent lattices. This region is also the region where temperature gradient should be expected. Note that in particular this merging method implies that the notion of "temperature" in the middle  of the system is questionable. On the other hand this approach seems the simplest one. It will be referred to as the tensor product  approach.

The data shown in Fig.~\ref{fig:fluct} -- as well as in all results of quasi-exact METTS simulations in the sequel of this paper -- display 
site-to-site fluctuations. This is an intrinsic drawback due to the statistical description of the system state in the METTS method; it also
affects all statistical methods \textit{\`a la Monte-Carlo.} These short range fluctuations decrease when the number of METTS states increases. Note also that
they decay in the course of the temporal evolution, as clearly seen in Fig.~\ref{fig:fluct}. 
They are responsible for the noisy character of some figures shown below, but they do not affect any of the
conclusion of this paper.

\subsubsection{Smooth Gluing}
\label{sec:smooth}

Another approach is possible which makes the transition region between {the} two subsystems more subtle. 
We prepare the initial state in just a single step: the inhomogeneous system containing the ``left'' and ``right'' parts with different temperatures is prepared in a single canonical ensemble simulation \emph{at thermal equilibrium}.
To achieve that, we consider an auxiliary 
Hamiltonian 
${\cal H}=\sum_i H_i f(i),$
with the scaling function $f$ smoothly interpolating between $\beta_L$ on the left side and $\beta_R$ on the right side. 
For example,
we have used:
\be \label{glue}
f(i)= \frac{\beta_R-\beta_L}{2} \tanh(i/s)+\frac{\beta_R+\beta_L}{2}
\ee
where $1\ll s\ll M$ denotes the size of the interface.
We then construct the canonical density matrix for the auxiliary
Hamiltonian at $\beta=1$.
Then, provided the correlation length in the system is much shorter than the interface size $s$ it is expected that 
the reduced density matrix of  the left part $\rho_L=\Tr_{R}\exp(-{\cal H}) \approx \exp(-\beta _L H_L),$ while that for  the right part reads $\Tr_{L}\exp(-{\cal H}) \approx \exp(-\beta _R H_R).$ 
This agreement requires also achieving a thermodynamic limit in terms of system size.

\subsection{Observables and thermometry}

The typical observables measured in the created ensemble or during the subsequent temporal evolution is the 
average number of particles per site $\av{n_i}$ as well as its variance $\Var(n_i)=\av{n_i^2}-\av{n_i}^2$. The variance at finite  
temperature contains both the quantum and the thermal fluctuations and can be used as a measure of the temperature. 
The advantage of this measure is that it is local, and thus well suited for the situation of interest where there
are a space-dependent temperature profile and heat currents. It is however \textit{a priori} not obvious that the local variance is a faithful measure of
the local temperature. In order to test this assumption (similar in spirit to the well-known local density approximation used for optical lattices
exposed to a slowly spatially varying trapping potential, see e.g. \cite{urba06})
we consider smoothly glued samples -- see section~\ref{sec:smooth}    
  with $M=35$ (at unit mean density: 70 bosons on 70 sites), $J/U=0.05$ and $\beta_L U=8, \beta_R U=12$. 

\begin{figure}
\includegraphics[width=8cm]{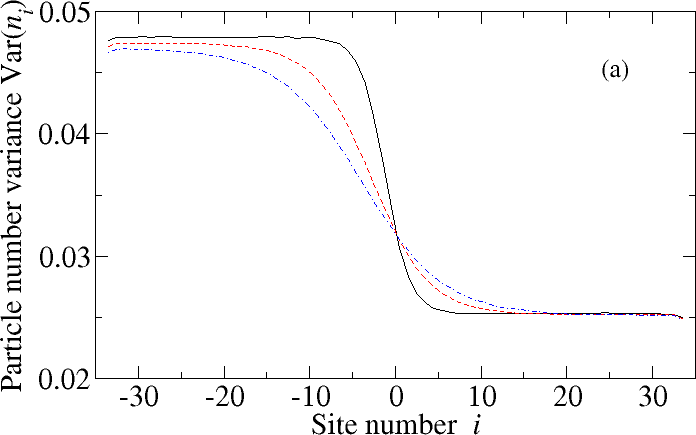}
\includegraphics[width=8cm]{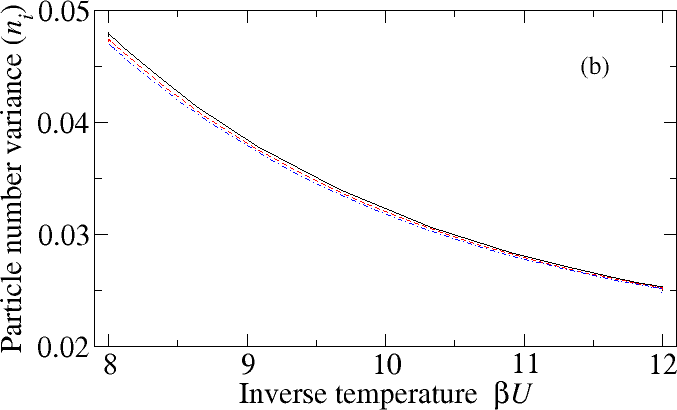}
\caption{(color online) Local thermometer based on the particle number variance. (a): local variance on the various sites for two
smoothly glued Mott insulators, with inverse temperatures $\beta_L U=8$ (left part) and $\beta_R U=12$ (right part). The temperature profile
is given by Eq.~(\ref{glue}), with a (smooth) interface size $s=3$ (black solid line), $s=7$ (red, dashed line) and $s=10$ (blue, dot-dashed line).
(b): local variance vs. the (inverse) local temperature, Eq.~(\ref{glue}). Except for small differences, the three curves collapse,
proving that the local variance -- with a good approximation -- depends only on the local temperature, regardless of the sharpness of the temperature profile. This proves that the local variance can be used as a reliable thermometer in a inhomogeneous system.
}
\label{term}
\end{figure}

The Quantum Monte Carlo (QMC) approach as implemented in ALPS~\cite{Bauer:Alps2:JSM11} allows to find the canonical ensemble density of ${\cal H}:$ $\rho\propto \exp(-{\cal H})$ and consequently to compute the
the variance $\Var(n_i)$ for a temperature inhomogeneous system. The function $f(i)$,  eq.~(\ref{glue}), is  the local (inverse) temperature.
In Fig.~\ref{term}, we show the local variance $\Var(n_i)$ vs. the local inverse temperature. Except for small differences related to finite size effects, all the data collapse on the same curve, regardless of the sharpness of the transition between the hot and the cold parts of the system.

\subsection{Real time evolution}

Real time evolution of each member of a METTS ensemble is performed using a home-made implementation of the TEBD algorithm, taking advantage of total particle number conservation  \cite{Zakrzewski09}.  
The ensemble consists typically of about 12000 METTS. By restricting to MPS of maximum bond dimension  $\chi=150,$ 
we have been able to perform numerical evolution up to time $tU=120,$ (corresponding to $tJ=6$) keeping discarded weights at the level of 
$10^{-4}$ at most.  We have used a fourth order Trotter decomposition in time to control the 
time discretization error \cite{Daley04}.
As we are able, see sections~\ref{sec:tp} and \ref{sec:smooth}, to prepare an initial state far from the thermal equilibrium,
and we can measure both its density and temperature profiles 
as a function of increasing time, we can {\sl directly} monitor heat
and mass transport in the system. We can moreover study quantitatively transport properties by calculating the energy and particle
currents in the system.
 We follow to some extent the approach presented in \cite{Karrasch13} for spin systems. 

Let us rewrite first the Bose-Hubbard system in a symmetrized form as
$H=\sum_i h_i$ with 

\bea \label{version-a}
h_i&=&-\frac{J}{2}\left(a_ia_{i+1}^\dagger+a_{i+1}a_{i}^\dagger+a_ia_{i-1}^\dagger +a_{i-1}a_{i}^\dagger \right)\nonumber\\
&& + \frac{U}{2} n_i(n_i-1).%-\mu n_i,
\eea
The currents flowing in and out of site $i$ may be defined via continuity equations. For example the energy current 
may be obtained from 
\bea
\partial_t h_i&=& i[H,h_i]\nonumber \\
&=& i[h_{i-2},h_i]+ i[h_{i-1},h_i]+ i[h_{i+1},h_i]+i[h_{i+2},h_i] \nonumber \\
&\equiv&  \jE(i-1/2)-\jE(i+1/2),
\eea
where $ \jE(i-1/2)$ is the current coming into site $i$ from site $i-1$. 
Since the site index $i$ is in our case a 
half-integer number, the current index is an integer. In particular $\jE(0)$ corresponds to a current passing through the center where the two subsystems are glued together.
A little algebra shows  that 
\be \label{curr}
\jE(i-1/2)= i[h_{i-2},h_i]+ i[h_{i-1},h_i]+i[h_{i-1},h_{i+1}],\ee
with the first and last terms  being the consequence of our symmetrized form of  $h_i$ in (\ref{version-a}) which involves a site energy plus half of the links to neighbors on both sides.

The Hamiltonian for short range interactions on a lattice may be split as a sum of site terms in several ways, leading to slightly
different expressions for the currents. For example, \cite{Karrasch13} uses an asymmetric site Hamiltonian containing a site energy and the full link to the right.
However, the different forms lead to very similar results whenever the system changes smoothly from site to site.

For our Bose-Hubbard system, the energy current operator defined in Eq.~(\ref{curr}) reads explicitly:
\bea \label{jEa}
\jE(i-1/2)&=\frac{J^2}{4}\left [ {\cal K}(i-2,i)+{\cal K}(i-1,i+1)\right]%+\mu J{\cal K}(i-1,i)
\nonumber \\
&+\frac{UJ}{2} \left [ {\cal N}(i,i-1) -  {\cal N}(i-1,i) \right ]
\eea
with  ${\cal K}(i,j)\equiv i( a^\da_i a_j - a^\da_j a_i)$  and $   {\cal N}(i,j)\equiv i(a^\da_i n_i a_j - h.c.) $ (with $n_i= a^\da_i a_i$). 

Similarly we may define the mass (particle) current $\jN(i=1/2)$ \footnote{Strictly speaking, we are defining a particle current. In order
to avoid any ambiguity with the quasiparticle/quasihole currents defined in Section~\ref{Sec:bogoliubov}, we prefer to use the
unambiguous words ''mass current''.} using the continuity equation for $\partial_t n_i$
\bea
\partial_t n_i&=& i[H,n_i]\nonumber \\
&=& i[h_{i-1},n_i]+ i[h_{i},n_i]+i[h_{i+1},n_i] \nonumber \\
&\equiv&  \jN(i-1/2)-\jN(i+1/2)
\eea
yielding 
\be \label{jN}
\jN(i-1/2)=J{\cal K}(i,i-1).
\ee

\begin{figure}
\includegraphics[width=8cm]{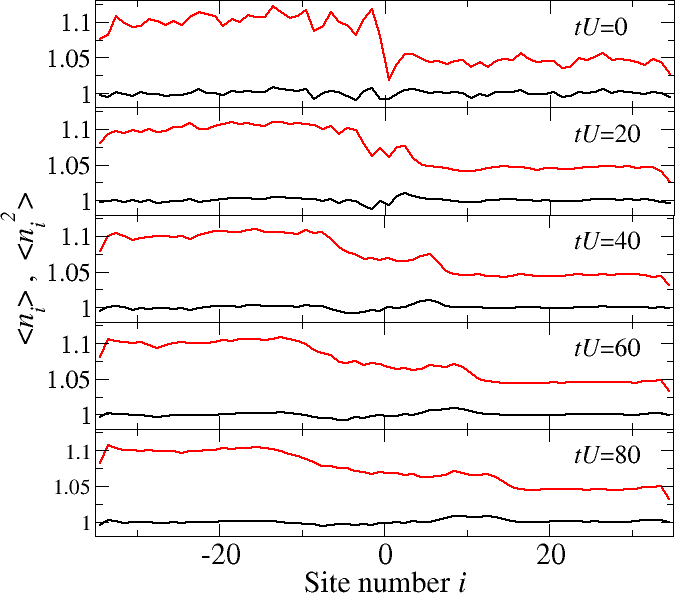}
\caption{(color online) Dynamical evolution of the average particle number $\av{n_i}$ and average squared particle number 
$\av{n_i^2}$  after two Mott insulators at different temperatures $\beta_L U=6,\beta_R U=8$ are sharply connected
at time $t=0.$ Each subsystem has initially 35 particles on 35 sites. Note the spreading 
of the central perturbation observed both on the average density $\av{n_i}$ (lower, black curves in all the panels) and on the average squared density $\av{n_i^2}$ (higher, red lines in all the panels).
}
\label{merge1}
\end{figure}

Having all the observables defined, let us take a closer look at exemplary numerical data. 
We consider two sharply connected Mott insulators (with $M=35$ each) at $\beta_L U=6$ and $\beta_R U=8$. 
Again the system contains 70 particles in total. For short times, the variance for this run has already been shown in Fig.~\ref{fig:fluct}. 
Fig.~\ref{merge1} shows  the average density, ${\av{n_i}},$ and the averaged squared density, ${\av{n_i^2}},$ for longer times.
Firstly observe that the site to site fluctuations
of both quantities are larger than on the variance $\av{n_i^2}-\av{n_i}^2$ (subtraction reduces fluctuations), 
compare with the lowest panel in Fig.~\ref{fig:fluct}).
Apart from statistical 
fluctuations one may clearly observe the spreading of the perturbation from the center of the sample with time. 
The density shows a {small but} clear excess above one on the right hand side (the cold one), this excess moves to the right with an apparently constant velocity.  
This excess must be somewhere compensated by a lack of particles (the particle number is conserved): 
indeed a hole moves to the left (hot side). 
Similarly bumps and holes in $\av{n_i^2}$ spread out in both directions approximately linearly in time.

\begin{figure}
\includegraphics[width=8cm]{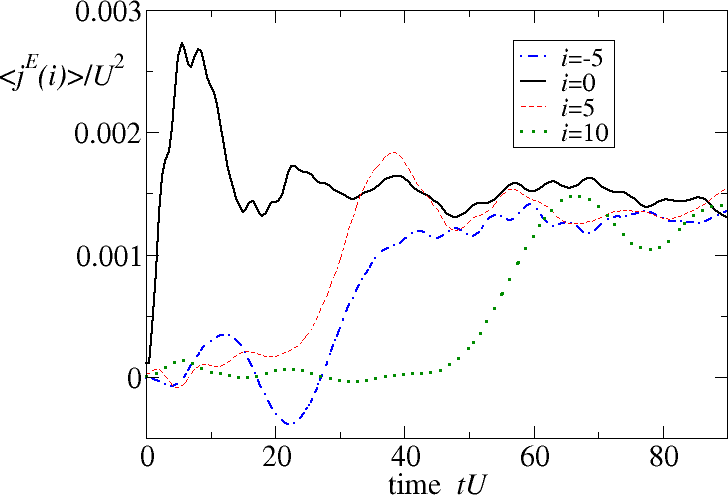}
\caption{(color online) Energy currents associated with the data of Fig.~\ref{merge1}. The current initially starts flowing at the connecting point, $i=0$, between
the two samples at different temperatures. It then propagates on both sides, reaching sites $i=\pm5$ around $tU=25,$ and site $i=10$ later around $tU=50.$
 }
\label{merge2}
\end{figure}

\begin{figure}
\includegraphics[width=8cm]{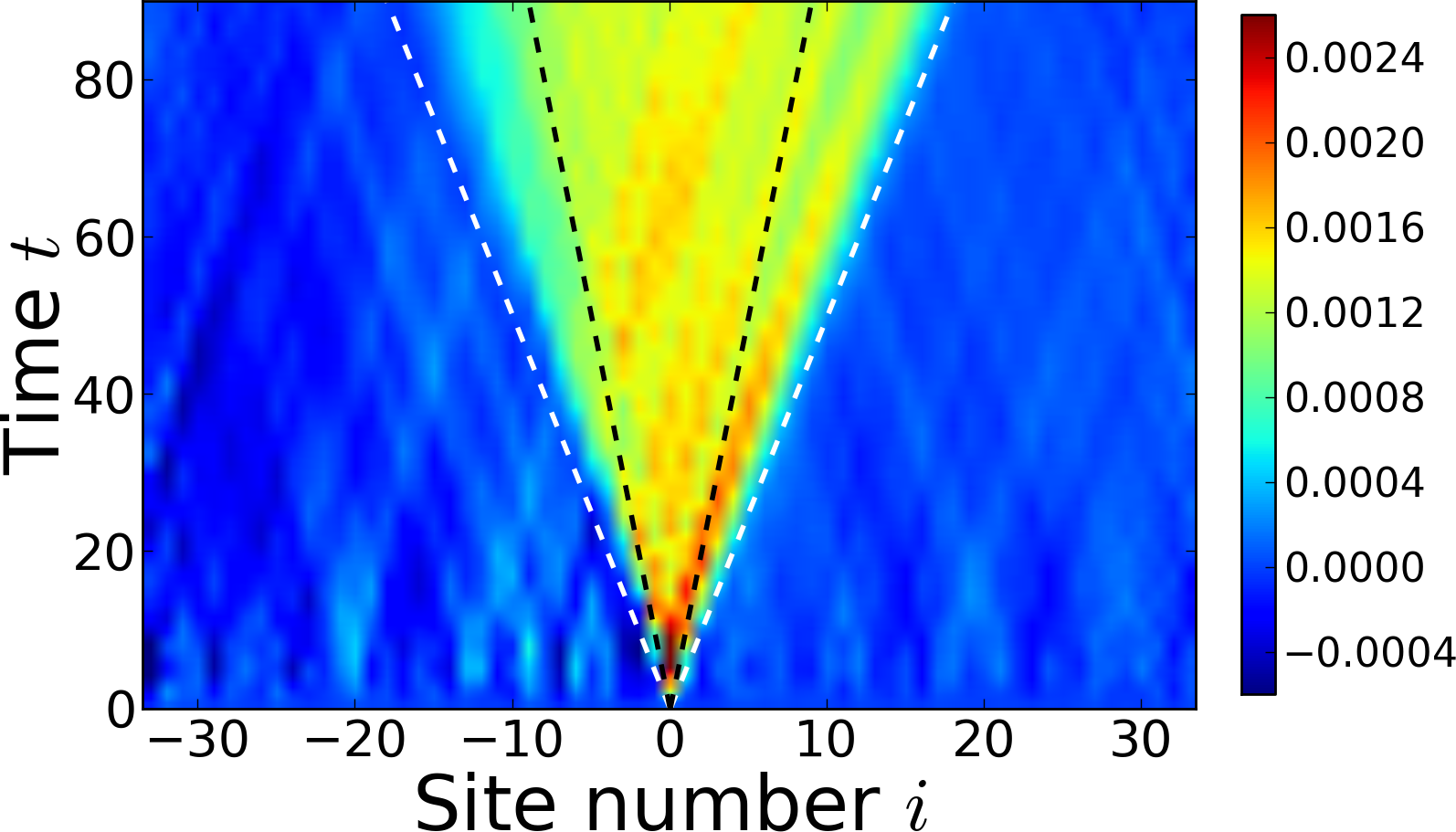}
\caption{(color online) Spatio-temporal development of the energy current  for the data 
 represented in Fig.~\ref{merge1}. 
Observe the ballistic spreading of the perturbation, initially born at the connecting
point,  {$i=0$}, between the two Mott insulators at different temperatures. See text for further discussion.}
\label{fig:3D1}
\end{figure}

This picture is confirmed by observing the energy currents shown in Fig.~\ref{merge2}. 
One expects that, when gluing two subsystems together, the effect appears first {around the gluing point}. 
Then the information travels within the system modifying
various observables (including currents). We observe this effect in Fig.~\ref{merge2}. The current at the place of merging, $\jE(0)$ almost immediately picks up non zero values starting from the gluing time $t=0$. 
A crucial observation is that, after some transient, the current $\jE(0)$ is constant at long time, although 
the spatial profile of $\avn$ and $\avnsq$ becomes smoother and smoother. This definitely rules out the
possibility of diffusive energy transport in the system. Indeed, in such a case, the current would have been proportional
to the local temperature gradient, and would decay at long times. The persistence of a finite current at long times is 
 a direct signature of ballistic transport.
This is confirmed by looking at the current at other sites. They fluctuate around zero at short time,
and rise only after some position-dependent delay.
The spread of the information is approximately linear in time (ballistic) as visible in Fig.~\ref{merge2} and more apparent while inspecting the 
color plots in Fig.~\ref{fig:3D1}. 
The dashed lines give the predicted theoretical limits -- derived in the next Section -- 
for the ballistic spreading of information. 

%%%%%%%%%%%%%%%%%%%%%%%%%%%%%%%%%%%%%%%%%%%%%%%%%%%%%%%%%%%%%%%%%%%%%%%%%%%%%%

\section{Bogoliubov theory}
\label{Sec:bogoliubov}
\subsection{Reduction of the Hilbert space}
For a moderate value of the tunneling amplitude $J,$ the Bose-Hubbard ground state at integer filling 
is a Mott insulator \cite{Mott}, with a non-zero gap. In order to understand heat and mass transport at finite
temperature, it is thus useful to study the properties of the ground state as well as those of the low energy
excitations. In this section, we give a complete quantitative description of these quantities,
in the limit of small tunneling $J\ll U$ and relatively 
{low temperature $\beta U \gg 1.$} Note that we
do not put any restriction on the parameter $\beta J,$ so that this approach takes into account
both the quantum fluctuations induced by $J,$ the thermal fluctuations governed by $\beta,$ and mixed contributions.
We show below that this simple theory is able to quantitatively reproduce the 
quasi-exact numerical results obtained for heat and mass transport in the fully out of equilibrium
Bose-Hubbard model.

In the limit of vanishing $J,$ the ground state at unit filling is simply a Fock state with exactly
one particle per site. Elementary excitations are particle-hole excitations where a second 
particle is added on one site while leaving another site empty. This costs an energy $U.$
If one drops the requirement of a fixed total number of particles -- thus going from a canonical to a grand canonical
ensemble -- elementary excitations are independent particle or hole excitations. The energy cost of the
particle excitation is $U-\mu$, where $\mu$ is the chemical potential, and the 
cost of a hole excitation is $\mu.$ The system is thus gaped for any value of $\mu$ in the $[0,U]$ interval.
At finite temperature, particles or holes can be created, with probabilities respectively $\rme^{-\beta(U-\mu)}$
and $\rme^{-\beta\mu}.$ The average density will be maintained at 1 if $\mu=U/2.$ Higher excitations (several particles
and/or holes) are negligible if $\beta U \ll 1.$

For a small but finite $J,$ the ground state is no longer a Fock state, but it remains gaped. 
Using a perturbative approach in powers of $J/U,$ it is possible~\cite{Damski} to obtain accurate estimates
of all interesting quantities such as energy, average occupation number $\avn$ or $\avnsq$ on site $i$.
We will here use a similar approach, restricting to lowest non-vanishing order in $J/U,$ but
including also elementary excitations. While particle/hole excitations on all sites are degenerate for $J=0,$
tunneling between sites lifts this degeneracy producing a band of quasiparticle and quasihole excitations.

At lowest order in $J/U,$ it is enough to consider single 
particle and hole excitations, that is to restrict the Hilbert space on each site
to occupation numbers $n_i=0,1,2.$ 
This problem has been addressed  in differents works,see e.g. \cite{Elstner99,Oosten01}; for a recent review with an extensive list of references, see \cite{Krutitsky15}. In  our description below,  following~\cite{Cucchietti07}, we
define a ''vacuum'' state as the Fock state
with one particle on each site and introduce
particle and hole creation operators on site $i:$ $\cd_i$ and $\dd_i$.
Their action is best illustrated by mapping the boson occupation number onto two quantum numbers $(n_c,n_d)$ 
representing particle and hole occupation numbers: on each site, we have $|2\rangle=|(1,0)\rangle$,
$|1\rangle=|(0,0)\rangle$ and $|0\rangle=|(0,1)\rangle$. One then has to implement that there cannot be less 
than zero or more than two bosons on each site and that states with one particle and one hole
on the same site are forbidden.

This results in an effective Hamiltonian quadratic in particle/hole operators, see~\cite{Cucchietti07}
for a detailed derivation:
\begin{equation}
H_{\mathrm{eff}} =  -J \sum_{\langle i,j\rangle}{\left[ 2\cd_i c_j + \dd_i d_j + \sqrt{2} (c_i d_j + \cd_i \dd_j)\right]}
+ U \sum_i{\cd_i c_i}
\end{equation}
where $\langle i,j\rangle$ denotes a pair of neighboring sites.

When the density of excitations is small, $\langle \cd_i c_i\rangle, \langle \cd_i c_i\rangle \ll 1,$
the $c_i,\cd_i,d_i,\dd_i$ can be approximately considered as standard bosonic annihilation and creation operators.

\subsection{Bogoliubov approach}
For a finite system with periodic boundary conditions (or an infinite system), this effective Hamiltonian can be 
diagonalized using the translational invariance, going to the momentum space \cite{Cucchietti07}:
\begin{equation}
c_k = \frac{1}{L} \sum_r{c_r \rme^{-\rmi kr}}\ \ ,\ \  d_k = \frac{1}{L} \sum_r{d_r \rme^{-\rmi kr}} 
\end{equation}
where $L$ is the number of sites of the system.
The effective Hamiltonian writes:
\bea
 H_{\mathrm{eff}} &=-2J \sum_k{\cos k \left[ 2 \cd_k c_k + \dd_k d_k + \sqrt{2} (c_k d_{-k} + \cd_k \dd_{-k}) \right]} \nonumber\\
 &+ U \sum_k{\cd_k c_k}
\eea
which can be conveniently rewritten as:
\begin{eqnarray}
H_{\mathrm{eff}}&=&\sum_k[\cd_k,-d_{-k}] 
\left[
\begin{array}{cc}
U-4J\cos k & -2\sqrt{2}J\cos k \\
2\sqrt{2}J\cos k  & 2J\cos k
\end{array}
\right]
\left[
\begin{array}{c}
c_k \\ \dd_{-k}
\end{array}
\right].\nonumber\\ 
\label{H22}
\end{eqnarray}
This Hamiltonian can be diagonalized using a Bogoliubov transformation:
\begin{equation}
 c_k = u_k B_k + v_k^* \Ad_{-k}\ \ ,\ \ \dd_{-k}=v_k B_k + u_k^*\Ad_{-k}
\end{equation}
where $(u_k,v_k)$ is the eigenmode of the Bogoliubov-de Gennes equations:
\begin{eqnarray}
E_k^{(\mathrm{qp})}\!
\left[
\begin{array}{c}
u_k \\ v_k 
\end{array}\right]\!=\!
\left[\begin{array}{cc}
U-4J\cos k  & -2\sqrt{2}J\cos k \\
 2\sqrt{2}J\cos k  &  2J\cos k 
\end{array}\right]\!
\left[
\begin{array}{c}
u_k \\ v_k 
\end{array}\right]
\label{Eq:Bogoliubov_eigenmodes}
\end{eqnarray}
with eigenenergy:
\begin{equation}
E_k^{(\mathrm{qp})} = \frac{U}{2}-J \cos k + \omega_k, 
\label{Eq:energy_qp}
\end{equation}
where
\begin{equation}
\omega_k = \frac{\sqrt{U^2-12JU\cos k+4J^2\cos^2 k}}{2}.
\end{equation}
With the normalization $|u_k|^2-|v_k|^2=1,$ the operators $A_k,B_k,\Ad_k,\Bd_k$ are standard 
bosonic annihilation and creation operators.
The other eigenmode $(-v_k,u_k)$ is associated with eigenenergy:
\begin{equation}
E_k^{(\mathrm{qh})} = -\frac{U}{2}+J \cos k + \omega_k,  
\label{Eq:energy_qh}
\end{equation}
so that the Hamiltonian writes:
\begin{equation}
H_{\mathrm{eff}} = \sum_k{\left[E_k^{(\mathrm{qp})}\Bd_k B_k + E_k^{(\mathrm{qh})}\Ad_k A_k + \omega_k + 3J \cos k - \frac{U}{2}\right]}. 
\label{Eq:H2}
\end{equation}
Eqs.(\ref{Eq:energy_qp}-\ref{Eq:energy_qh}) are special cases of general formulae avaliable for arbitrary integer mean density (see e.g. \cite{Oosten01}) as reviewed by Krutisky \cite{Krutitsky15}, see his Eq.~(325).

\subsection{Ground state}
The ground state of the system is the vacuum of the $A_k,B_k$ operators and its energy is $\sum_k{(\omega_k + 3J \cos k - U/2)}.$
In a finite system with periodic boundary conditions, the $k$ values are discretized as integer multiples of $2\pi/L.$ 
The limit of large system is obtained with the substitution:
\begin{equation}
 \frac{1}{L} \sum_{k} \to \int_{-\pi}^{+\pi} \frac{\rmd k}{2\pi}.
\end{equation}
In the limit of small $J$ we are interested in, one gets:
\begin{equation}
 \omega_k \approx \frac{U}{2} - 3J\cos k - 8 \frac{J^2}{U} \cos^2 k
 \label{Eq:omegak}
\end{equation}
so that the ground state energy is approximately equal to $-4J^2 L/U,$ ($-4J^2/U$ per site), in agreement with \cite{Damski}
(note that the first order in $J$ cancels out).

It is also possible to compute expectation values of simple operators in the ground state. The total number of particles is:
\begin{equation}
 N = \sum_i{n_i} = \sum_i{(1+ \cd_i c_i - \dd_i d_i)} = L+ \sum_k{\cd_k c_k}.
\end{equation}
It is simply expressed as:
\begin{equation}
 N = L + \sum_k{(\Bd_k B_k - \Ad_k A_k)}
 \label{Eq:N}
\end{equation}
so that $\Bd_k$ (resp. $\Ad_k$) appears as the creation operator for a quasiparticle (resp. quasihole) with momentum $k$.
In the ground state, one trivially recovers $\langle N\rangle = L,$ so that there is on average exactly one particle per site.

The variance of the number of particle per site is more interesting. Indeed:
\begin{equation}
 N_2 = \sum_i{n_i^2} =  \sum_i{(1+ \cd_i c_i - \dd_i d_i)^2},
\end{equation}
which gives, in the limit of low density of excitations:
\begin{equation}
 N_2 = L + \sum_i{(3 \cd_i c_i - \dd_i d_i)} = L + \sum_k{(3 \cd_k c_k - \dd_k d_k)}.
\end{equation}
The physical interpretation of the +3 and -1 coefficients is rather transparent: 
 $\cd_i$ creates an additional particle at site $i,$
thus $n_i^2=4,$ i.e. 3 on top of the 1 particle of the background. Similarly, $\dd_i$ produces $n_i=0$, that
is 1 below the background.

In turn, the operator $N_2$ can be expressed as a function of the $A_k$ and $B_k$ operators:
\bea
 N_2 =& L + \sum_k{\left[(3|u_k|^2-|v_k|^2)\Bd_k B_k - (|u_k|^2-3|v_k|^2)\Ad_k A_k \right.} \nonumber\\
 & \left.+ 2 u_k v_k A_k B_k + 2 u_k^* v_k^* \Ad_k
 \Bd_k + 2 |v_k|^2 \right].
 \label{Eq:N2}
\eea
In the limit of small $J,$ the eigenmodes of Eq.~(\ref{Eq:Bogoliubov_eigenmodes}) are:
\begin{equation}
 u_k=1\ \ ,\ \ v_k=2\sqrt{2} \cos k\ J/U,
\end{equation}
which, in a more general form can be found also in  \cite{Krutitsky15}, see  Eq.~(326) there.
The expectation values on the ground state is:
\begin{equation}
 \langle N_2 \rangle = L +\sum_k{16 \cos^2 k \frac{J^2}{U^2}} = L \left(1+\frac{8 J^2}{U^2}\right), 
 \label{Eq:N2_T0}
\end{equation}
recovering the fact that the variance of the number of particle per site is $8 J^2/U^2$ \cite{Damski}.

\subsection{Thermal excitations}
\label{Sec:bogo_thermal}
We now turn to thermal excitations in the system. One can
use either the canonical or the grand canonical ensemble, all expectation values are expected to be equal in the two ensembles in 
the large $L$ limit. It turns out that the grand canonical ensemble is more convenient for calculations.
We thus have to consider configurations with a weight scaling like $\rme^{-\beta(H_{\mathrm{eff}}-\mu N)}.$
The operator $H_{\mathrm{eff}}-\mu N$ is readily obtained from Eqs.~(\ref{Eq:H2}),(\ref{Eq:N}):
\bea
H_{\mathrm{eff}} - \mu N &= \sum_k{\left[(E_k^{(\mathrm{qp})}-\mu)\Bd_k B_k + (E_k^{(\mathrm{qh})}+\mu) \Ad_k A_k \right.}\nonumber\\
&\left.+ \omega_k + 3J \cos k - \frac{U}{2}\right]. 
\label{Eq:H2_GC}
\eea
The calculation of expectation values at the thermal equilibrium is thus straightforward. One obtains
$\langle B_k\rangle = \langle \Bd_k\rangle= \langle A_k\rangle= \langle \Ad_k\rangle = 0$ and:
\begin{equation}
 \langle \Bd_k B_k \rangle = \frac{\Tr\left[\Bd_k B_k \rme^{-\beta(E_k^{(\mathrm{qp})}-\mu)\Bd_k B_k}\right]}{\Tr\left[\rme^{-\beta(E_k^{(\mathrm{qp})}-\mu)\Bd_k B_k}\right]},
\end{equation}
which, in the limit of low density of excitations, reduces to:
\begin{equation}
 \langle \Bd_k B_k \rangle = \rme^{-\beta(E_k^{(\mathrm{qp})}-\mu)}.
 \label{Eq:NB}
\end{equation}
Similarly:
\begin{equation}
 \langle \Ad_k A_k \rangle = \rme^{-\beta(E_k^{(\mathrm{qh})}+\mu)}.
\label{Eq:NA}
 \end{equation}
We thus obtain for the average number of particles:
\begin{equation}
 \langle N \rangle = L +\sum_k{\left[\rme^{-\beta(E_k^{(\mathrm{qp})}-\mu)} - \rme^{-\beta(E_k^{(\mathrm{qh})}+\mu)} \right]}.
\end{equation}
In the limit of small $J/U,$ one can use the approximate expression (\ref{Eq:omegak}) at first order and obtain:
\begin{equation}
 \langle N \rangle = L +\sum_k{\left[\rme^{-\beta(U-4J\cos k-\mu)} - \rme^{-\beta(\mu-2J\cos k)} \right]}.
\end{equation}
In the continuous limit of large $L,$ the sum over $k$ becomes an integral giving:
\begin{equation}
  \frac{\av{N}}{L} = 1 + \rme^{-\beta(U-\mu)} \bessel(4\beta J) - \rme^{-\beta\mu} \bessel(2\beta J),
\end{equation}
where $\bessel$ is the modified Bessel function.

The physical interpretation of these equations is simple. At equilibrium, the system has a finite
density of quasiparticle $\nqp(k)$ and quasihole $\nqh(k)$ excitations, depending on the momentum $k$ and simply given
by Eqs.~(\ref{Eq:NB},\ref{Eq:NA}) which give in the small $J/U$ limit:
\begin{eqnarray}
 \nqp(k) & = & \rme^{-\beta(U-4J\cos k-\mu)} \nonumber\\
 \nqh(k) & = & \rme^{-\beta(\mu-2J\cos k)}, 
 \label{Eq:nqpqh}
\end{eqnarray}
so that one simply has:
\begin{equation}
\frac{\av{N}}{L} = 1+ \int_{-\pi}^{+\pi}{[\nqp(k) - \nqh(k)]\ \rmd k/2\pi}.
\label{Eq:avN}
\end{equation}

The chemical potential $\mu$ that ensures unit filling is thus obtained from the $\langle N \rangle = L$ constraint:
\begin{equation}
 \mu = \frac{U}{2} + \frac{1}{2\beta} \ln \frac{\bessel(2\beta J)}{\bessel(4\beta J)}.
 \label{Eq:mu_eq}
\end{equation}

Similarly, one can easily compute the variance of the local occupation number, relying on Eqs.~(\ref{Eq:N2}),(\ref{Eq:NB}),(\ref{Eq:NA}):
At lowest order in $J/U,$ we obtain:
\begin{equation}
\frac{\av{N_2}}{L} = 1 + 8 \frac{J^2}{U^2} + 3 \rme^{-\beta(U-\mu)} \bessel(4\beta J) - \rme^{-\beta\mu} \bessel(2\beta J),
\end{equation}
which, for the value of $\mu$ at unit average filling, gives:
\begin{equation}
 \frac{\av{N_2}}{L} = 1 + 8 \frac{J^2}{U^2} + 2 \rme^{-\beta U/2} \sqrt{\bessel(4\beta J) \bessel(2\beta J)}.
\label{Eq:N2_final}
\end{equation}
In the limit of zero-temperature, one recovers Eq.~(\ref{Eq:N2_T0}) with only quantum fluctuations due to $J.$
In the limit of vanishing $J,$ the modified Bessel functions tend to unity, and one has purely thermal fluctuations.
In the presence of both quantum and thermal fluctuations, the variance is not perfectly additive, 
because of crossed thermal-quantum effects.

Again, the interpretation in terms of density of quasiparticles and quasiholes is simple as:
\begin{equation}
\frac{\av{N_2}}{L} = 1 + 8 \frac{J^2}{U^2} + \int_{-\pi}^{+\pi}{[3\nqp(k) - \nqh(k)]\ \frac{\rmd k}{2\pi}}.
\label{Eq:avN2}
\end{equation}

The variance is finally obtained by combining this equation with Eq.~(\ref{Eq:avN}):
\bea
 \Var{(n_i)} & = & \av{n_i^2} - \av{n_i}^2 \nonumber \\
             & = & 8 \frac{J^2}{U^2} + \int_{-\pi}^{+\pi}{[\nqp(k) + \nqh(k)]\ \frac{\rmd k}{2\pi}}.
\eea
which shows that quasiparticles and quasiholes equally contribute to the increase of the variance.

Other expectation values can be computed as well. For example, the energy density is the expectation value of 
the operator $H_i,$ Eq.~(\ref{eqn:BH}), and is given by:
\begin{equation}
 \av{H_i} = -4 \frac{J^2}{U} +  U \int_{-\pi}^{+\pi}{\nqp(k)\ \frac{\rmd k}{2\pi}}.
\end{equation}

It must be emphasized that all these results are valid only when the density of excitations is low, $\nqp(k),\nqh(k)\ll 1,$ i.e.
when both $\av{n_i}$ and $\av{n_i^2}$ are close to unity.

We have checked using DMRG and Quantum Monte Carlo numerical simulations with ALPS~\cite{Bauer:Alps2:JSM11} the validity
of Eq.~(\ref{Eq:N2_final}), both in the grand canonical ensemble and in the canonical ensemble for large systems.
For example, for $J=0.05U,$ the prediction is that the variance of the occupation number at zero temperature
should be 0.02 per site, while the numerical result is 0.0198453.
The situation is a bit more complicated for thermal excitations, for two independent reasons:
\begin{itemize}
 \item Although $J/U=0.05$ seems  a rather small value, 
the numerical prefactors in e.g. Eq.~(\ref{Eq:omegak}) are such that one has to 
 compare $3J$ with $U/2.$ For $J/U>3/2-\sqrt{2}\approx 0.086,$ 
 the Bogoliubov eigenmodes become unstable and 
 the whole perturbative approach breaks down. Thus, even at $J/U=0.05,$ significant corrections are expected for the thermal fluctuations.
 For $\beta U=10,$ we find the variance of the occupation number to be 0.0429 while the prediction is 0.0385, a 10\% difference.
 At smaller $J/U,$ the relative difference is smaller.
 \item Finite size effects. We observed that the variance obtained using the canonical ensemble in a system of size $L$ has a rather
 strong dependence on $L,$ much larger than for the grand canonical ensemble, at least for small $J$ and low temperature. We believe that, when
 working in the canonical ensemble, fluctuations in the local occupation number are due to particles or holes jumping from other sites.
 If the number of sites is very large, they can provide at no cost the additional particles or holes. In contrast, if the number of sites
 is too small to provide these additional particles, the fluctuations and thus the variance will be reduced. As a rule of thumb,
 this effect is important when the product of the number of sites $L$ by the variance of the occupation number is not much larger than unity.
 For $L=100, J/U=0.05, \beta U=10,$ the variance is reduced to 0.0327.
\end{itemize}
In order to take into account these effects, a simple method is to slightly reduce the density of the quasiparticle
and quasihole excitations, Eq.~(\ref{Eq:nqpqh}), by multiplying them by a constant factor (independent of $k$)
slightly smaller than unity, which reduces the thermal fluctuations, leaving the quantum fluctuations unaffected.
This reduction factor is chosen to reproduce the initial number variance (at {$t=0$}) in each sub-sample: for the ''sharp gluing'' scenario where the right and left sub-samples are initially not connected (see e.g. Fig.~{\ref{fig:n_n2_time_60}}), two different reduction factors are used. For the ''smooth gluing'' scenario with an initial inhomogeneous temperature profile (see Fig.~{\ref{varmetts}}), thermal fluctuations can be globally provided by both sub-samples and a single reduction factor is consequently used.

\subsection{Transport in non-equilibrium systems}
We now consider the situation of systems, not at the thermal equilibrium, where we want to compute the
temporal evolution of expectation values of local operators as well as the currents flowing inside the system.
This is a very complicated problem in general; we will restrict to the Bose-Hubbard model within the simplified assumption discussed above, namely when only local occupation numbers 0, 1 and 2 are allowed on each site.
If we further assume that the system is everywhere close to the Mott insulator state with unit filling, so that the
density of excitations is small, we can use the previous Bogoliubov description in terms of quasiparticles and quasiholes. Note that this \emph{does not} require the system to be locally at a thermal equilibrium.

The Bogoliubov theory described in the previous sections  explicitly uses the translational invariance in order to obtain uncoupled elementary excitations with a well defined momentum. This is well suited to describe how excitations propagate, as it boils down to the dispersion relation of the excitations, Eqs.~(\ref{Eq:energy_qp},\ref{Eq:energy_qh}). For a system where translational invariance is broken by say a temperature gradient, this is less convenient, as one needs to build ''wavepackets'' coherently superimposing excitations with various $k.$
A mixed position-momentum representation is in such a case more convenient, similar to the Wigner phase-space
representation used to describe an ordinary quantum particle obeying the Schr\"odinger equation~\cite{Wigner}.
The temporal evolution of the Wigner representation is well approximated by the classical dynamics in the semiclassical
limit where the wavelength is much shorter than the typical size over which the Wigner function varies.
In our case, the typical wavelength of Bogoliubov excitations is $2\pi/k,$ of the order of the lattice spacing.
Thus, provided we consider a spatially smooth profile of temperature/excitations, we can use a classical
description of the system in terms of densities of quasiparticles/quasiholes $\nqp/\nqh$ depending on both the position $x$ (a continuous variable in this approximation) and the momentum $k,$ which obey the classical Liouville equations of evolution under the Hamiltonian $H_{\mathrm{eff}}$:
\begin{eqnarray}
 \frac{\partial \nqp(x,k,t)}{\partial t} = - \frac{\rmd E_k^{(\mathrm{qp})}}{\rmd k}\ \frac{\partial \nqp(x,k,t)}{\partial x} \nonumber\\
  \frac{\partial \nqh(x,k,t)}{\partial t} = - \frac{\rmd E_k^{(\mathrm{qh})}}{\rmd k}\ \frac{\partial \nqh(x,k,t)}{\partial x}
 \label{Eq:evolution_liouville} 
\end{eqnarray}
As various $k$ components do not interact, the evolution is straightforward:
\begin{eqnarray}
 \nqp(x,k,t)& = &\nqp\left
 (x-\vqp_k t,k,0\right)\nonumber\\
 \nqh(x,k,t)& = &\nqh\left(x-\vqh_k t,k,0\right) 
 \label{Eq:propagation}
\end{eqnarray}
where the velocities of quasiparticles and quasiholes are given by:
\begin{eqnarray}
 \vqp_k& = &\frac{\rmd E_k^{(\mathrm{qp})}}{\rmd k} \nonumber\\
 \vqh_k& = &\frac{\rmd E_k^{(\mathrm{qh})}}{\rmd k}. 
\end{eqnarray}
In the limit of small $J,$ they are{, see Eqs.~(\ref{Eq:energy_qp}) and (\ref{Eq:energy_qh})}:
\begin{eqnarray}
 \vqp_k& = &4J \sin k \nonumber\\
 \vqh_k& = &2J \sin k, 
\end{eqnarray}
which means that quasiparticles propagate twice faster than quasiholes.

A key and non-trivial point of this approach is that it predicts that quasiparticle/quasiholes excitations propagate \emph{ballistically}, 
as suggested by the quasi-exact numerical results using METTS presented above.

Eqs.~(\ref{Eq:propagation}) make it trivial to compute the total density of quasiparticles and quasiholes at a given
position:
\begin{eqnarray}
\nqp(x,t)& = & \int_{-\pi}^{+\pi}\frac{\rmd k}{2\pi}\ \nqp\left(x-\vqp_kt,k,0\right)\nonumber\\
\nqh(x,t)& = & \int_{-\pi}^{+\pi} \frac{\rmd k}{2\pi}\ \nqh\left(x-\vqh_kt,k,0\right)
\label{Eq:nqpqht}
\end{eqnarray}
From these formula, one can deduce the expectation values of any local observable, following the derivation in section~\ref{Sec:bogo_thermal}:
\begin{eqnarray}
 \av{n(x)}   & = & 1 +                     \nqp(x,t) - \nqh(x,t) \nonumber\\
 \av{n^2(x)} & = & 1 + 8\frac{J^2}{U^2} + 3\nqp(x,t) - \nqh(x,t) \nonumber\\
 \Var{(n(x))}   & = &     8\frac{J^2}{U^2} +  \nqp(x,t) + \nqh(x,t) \nonumber\\
 \av{H(x)}/U & = &    -4\frac{J^2}{U^2} +  \nqp(x,t) 
 \label{Eq:occupation_prediction}
\end{eqnarray}

Although the quasiparticle/quasiholes densities are not in principle directly observable, these equations show
that it is enough to measure two independent quantities, for example the average local occupation number and its variance to extract both the quasiparticle and the quasihole densities.

\bigskip

The Bogoliubov approach also makes it possible to compute the currents. From Eq.~(\ref{Eq:evolution_liouville}), it follows that one can define
quasiparticle/quasihole currents:
\begin{eqnarray}
 \jqp(x,k,t) & = & \vqp_k \ \nqp(x,k,t) \nonumber\\
 \jqh(x,k,t) & = & \vqh_k \ \nqh(x,k,t)
\end{eqnarray}
which automatically satisfy the continuity equation $\partial n/\partial t + \partial j/\partial x = 0.$

The current at a given position is simply summed over all momenta contributions:
\begin{eqnarray}
 \jqp(x,t) & \equiv & \int_{-\pi}^{+\pi}{\vqp_k\ \nqp(x,k,t)\ \frac{\rmd k}{2\pi}} \nonumber\\
 \jqh(x,t) & \equiv & \int_{-\pi}^{+\pi}{\vqh_k\ \nqh(x,k,t)\ \frac{\rmd k}{2\pi}}
\end{eqnarray}

Following Eqs.~(\ref{Eq:occupation_prediction}), the mass and energy currents are thus given by:
\begin{eqnarray}
\label{eqcur}
 \jn(x,t)   & = & \jqp(x,t) - \jqh(x,t)\nonumber \\
 \jE(x,t)/U & = & \jqp(x,t).
\end{eqnarray}
The physical interpretation of these equations is clear: in the small $J$ limit, the dominant contribution 
to energy is the interaction brought by sites with double occupation, associated with quasiparticle excitations. 
In contrast, a quasihole does not lead to a change of local energy and thus does not contribute
to the energy current. As expected, quasiparticles and quasiholes both contribute to the mass (density) current,
with opposite signs.

The previous set of equations describe the ballistic transport of quasiparticles and quasiholes
in the system. In general, the momentum distribution of quasiparticles and quasiholes at a given position is not 
given by a thermal distribution, Eq.~(\ref{Eq:nqpqh}). This implies that our description goes beyond a local
thermal equilibrium. 
Even if the initial state at $t=0$ is in a local thermal equilibrium described by Eq.~(\ref{Eq:nqpqh}), for
space dependent (inverse) temperature $\beta(x)$ and chemical potential $\mu(x)$, this property is lost during time
evolution.

While the integrals involved in Eqs.~(\ref{Eq:nqpqht}) have a trivial structure and are easily
numerically computed for arbitrary initial distributions, it is in general difficult to perform the integrals
analytically. There are however simple cases where it is possible. 

Let us first consider the situation where two half blocks, each at thermal equilibrium with unit filling, but with different temperatures $(\beta_L$,$\beta_R)$ (and consequently
different chemical potentials given by Eq.~(\ref{Eq:mu_eq})) are connected at time $t=0.$ Taking $x=0$ as the connection point, this implies that
the initial quasiparticle/quasihole distributions are given by:
\begin{eqnarray}
 \nqp(x,k,0)  = & \rme^{-\beta_L(U-4J\cos k-\mu_L)} \ \ & \mathrm{for}\ x<0,\nonumber\\
 \nqp(x,k,0)  = & \rme^{-\beta_R(U-4J\cos k-\mu_R)} \ \ & \mathrm{for}\ x>0,\nonumber\\
 \nqh(x,k,0)  = & \rme^{-\beta_L(\mu_L-2J\cos k)}   \ \ & \mathrm{for}\ x<0,\nonumber\\
 \nqh(x,k,0)  = & \rme^{-\beta_R(\mu_R-2J\cos k)}   \ \ & \mathrm{for}\ x>0.
\end{eqnarray}
The temporal evolution of each $k$ component is trivial: it consists of a step function moving at a velocity
$\vqp_k$ (resp. $\vqh_k$) for quasiparticles (resp. quasiholes). 
In effect the densities have a very simple geometrical structure: they are ''smoothed'' steps connecting
the asymptotic ''left'' density (on the left side) to the asymptotic ''right'' density (on the right side).
This step function is centered around the origin, keeps the same shape at any time, being simply stretched along the $x$-axis proportionally to time. Because quasiparticles (resp. quasiholes) have a maximum velocity $4J$ (resp. $2J$), the smooth step extends only inside a ''light cone'', in a finite range $[-4Jt,4Jt]$ (resp.  $[-2Jt,2Jt]$).
The situation is similar for the currents which vanish outside the light cones and keep the same
spatial structure, simply stretching linearly in time.

We could not perform analytically the integral over $k$ for the
density, but could evaluate it for the currents:
\begin{eqnarray}
 \jqp(x,t) & = &\frac{1}{\pi} \left[ \rme^{-\beta_L(U-\mu_L)} \frac{\sinh \beta_Lu}{\beta_L} -  \rme^{-\beta_R(U-\mu_R)} \frac{\sinh \beta_Ru}{\beta_R}\right ]\nonumber\\
 \jqh(x,t) & = &\frac{1}{\pi} \left[ \rme^{-\beta_L\mu_L} \frac{\sinh \beta_Lv}{\beta_L} -  \rme^{-\beta_R\mu_R} \frac{\sinh \beta_Rv}{\beta_R}\right ]
\end{eqnarray}
where $u=\sqrt{16J^2-x^2/t^2}, v=\sqrt{4J^2-x^2/t^2}.$ These expressions are valid inside the ''light cones''
$|x|\leq 4Jt$ (resp. $|x|\leq 2Jt$) for the quasiparticles (resp. quasiholes){; outside the light cones, the currents vanish.} 

The currents at the origin $x=0$ are especially simple. They do not depend on time and are given:
\begin{eqnarray}
 \jqp(0) & = &\frac{1}{\pi} \left[ \rme^{-\beta_L(U-\mu_L)} \frac{\sinh 4\beta_L J}{\beta_L} -  \rme^{-\beta_R(U-\mu_R)} \frac{\sinh 4\beta_R J}{\beta_R}\right ]\nonumber\\
 \jqh(0) & = &\frac{1}{\pi} \left[ \rme^{-\beta_L\mu_L} \frac{\sinh 2\beta_L J}{\beta_L} -  \rme^{-\beta_R\mu_R} \frac{\sinh 2\beta_R J}{\beta_R}\right ]
\end{eqnarray}
The fact that they are time-independent is a signature of ballistic propagation of the excitations, and nicely fits the observations
made in Fig.~\ref{merge2} on our quasi-exact numerical METTS simulations. Both currents are simply of the form $g(\beta_L)-g(\beta_R)$ for some function $g,$
a property already emphasized as a signature of ballistic transport in~\cite{Karrasch13}. In contrast, diffusive transport would be characterized
by currents depending on the local gradients.

In the regime of intermediate temperature $J \ll \frac{1}{\beta_R},\frac{1}{\beta_R} \ll U,$ the expression simplifies and the currents
have a ''semi-circle'' spatial shape:
\begin{eqnarray}
 \jqp(x,t) & \propto & \sqrt{16J^2-x^2/t^2}\nonumber\\
 \jqh(x,t) & \propto & \sqrt{4J^2-x^2/t^2}.
\label{eq:semi-circle}
\end{eqnarray}
At lower temperature ($\beta J$ of the order of unity or smaller), the shape is qualitatively similar.

\section{Comparison of the Bogoliubov theory with numerical data}

\begin{figure*}
\includegraphics[width=8cm]{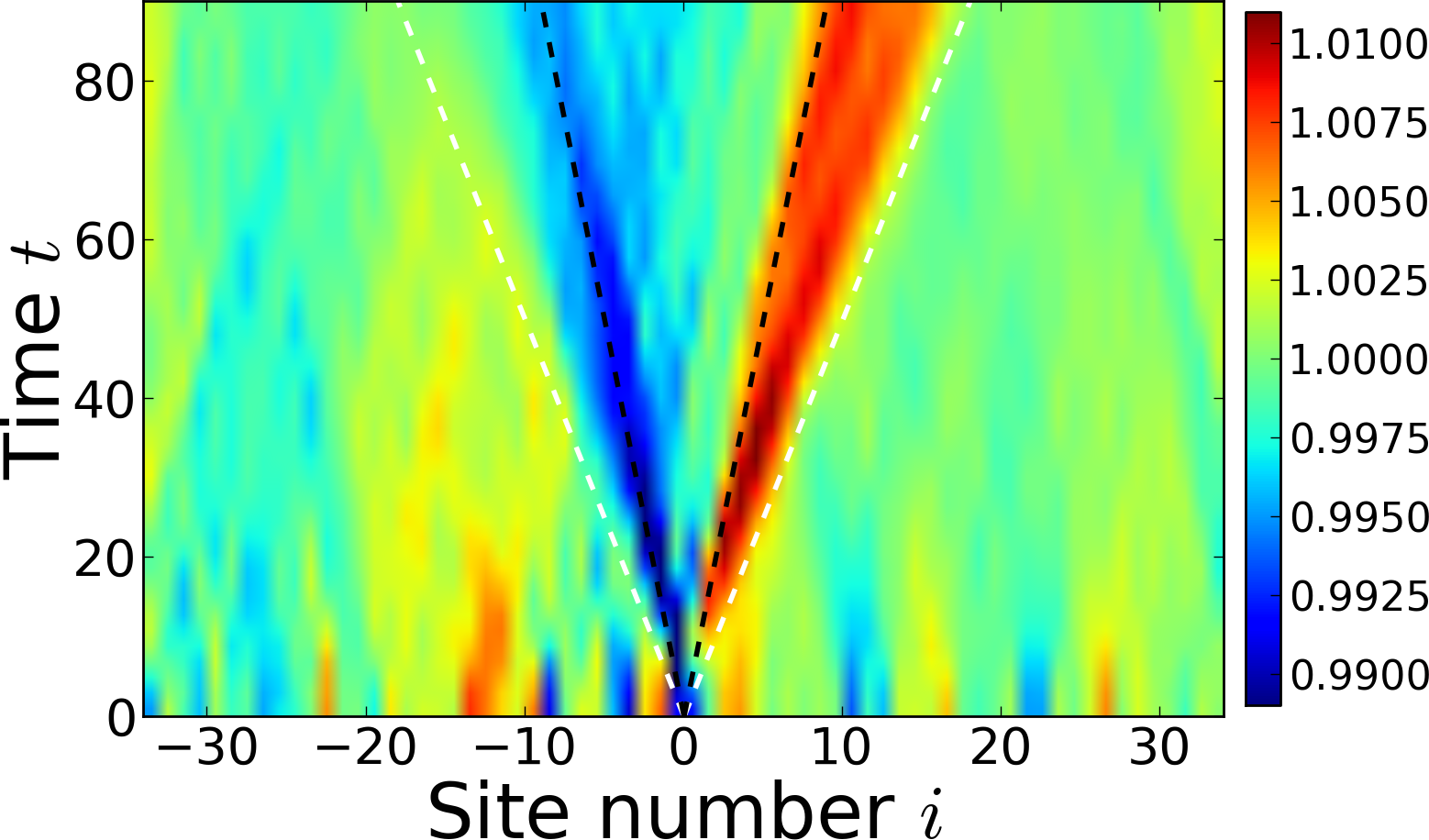}\vspace{2mm}\parbox[b][4.5cm][t]{1.6cm}{\huge$\langle n_i\rangle$}\includegraphics[width=8cm]{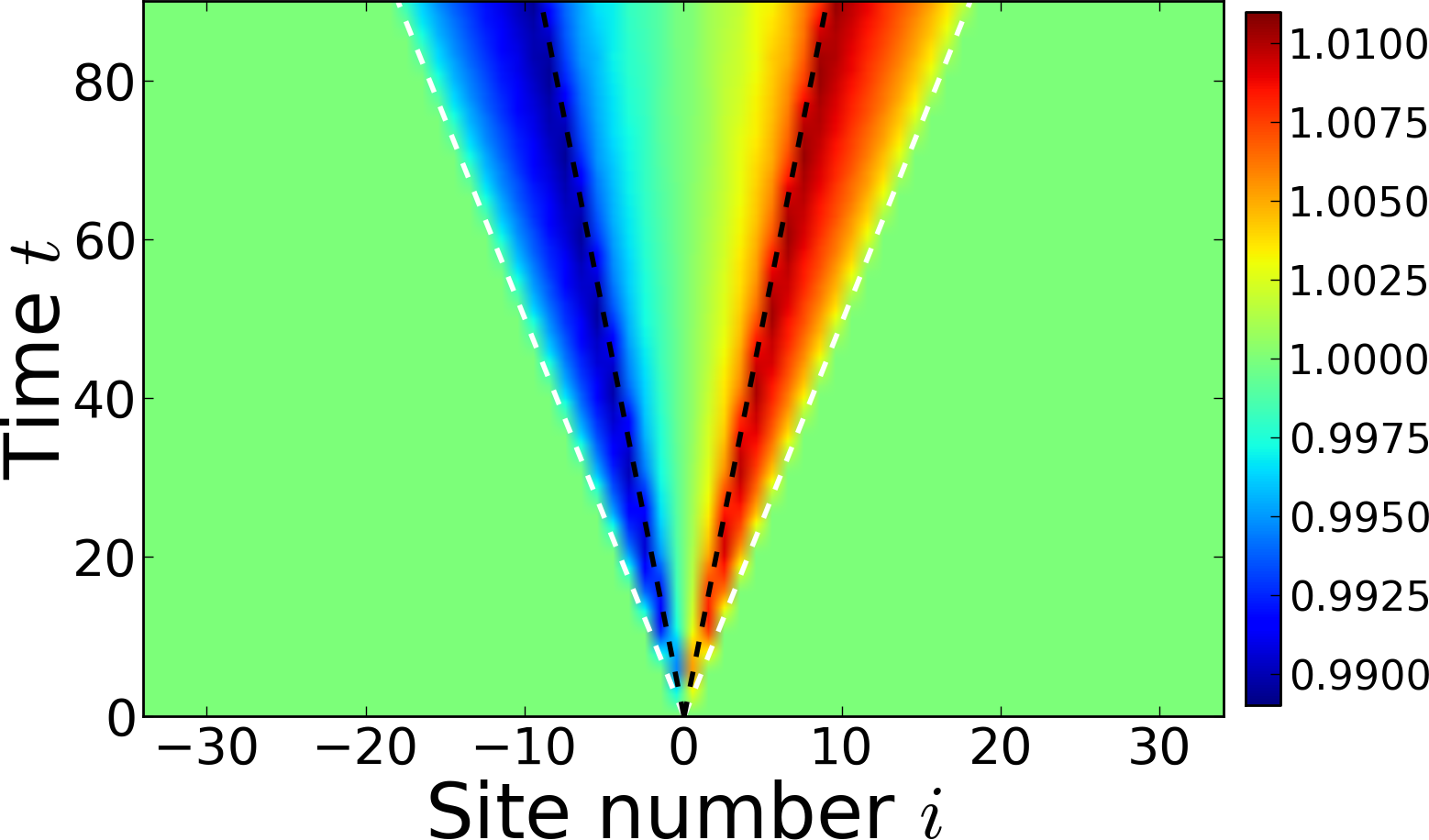}
\includegraphics[width=8cm]{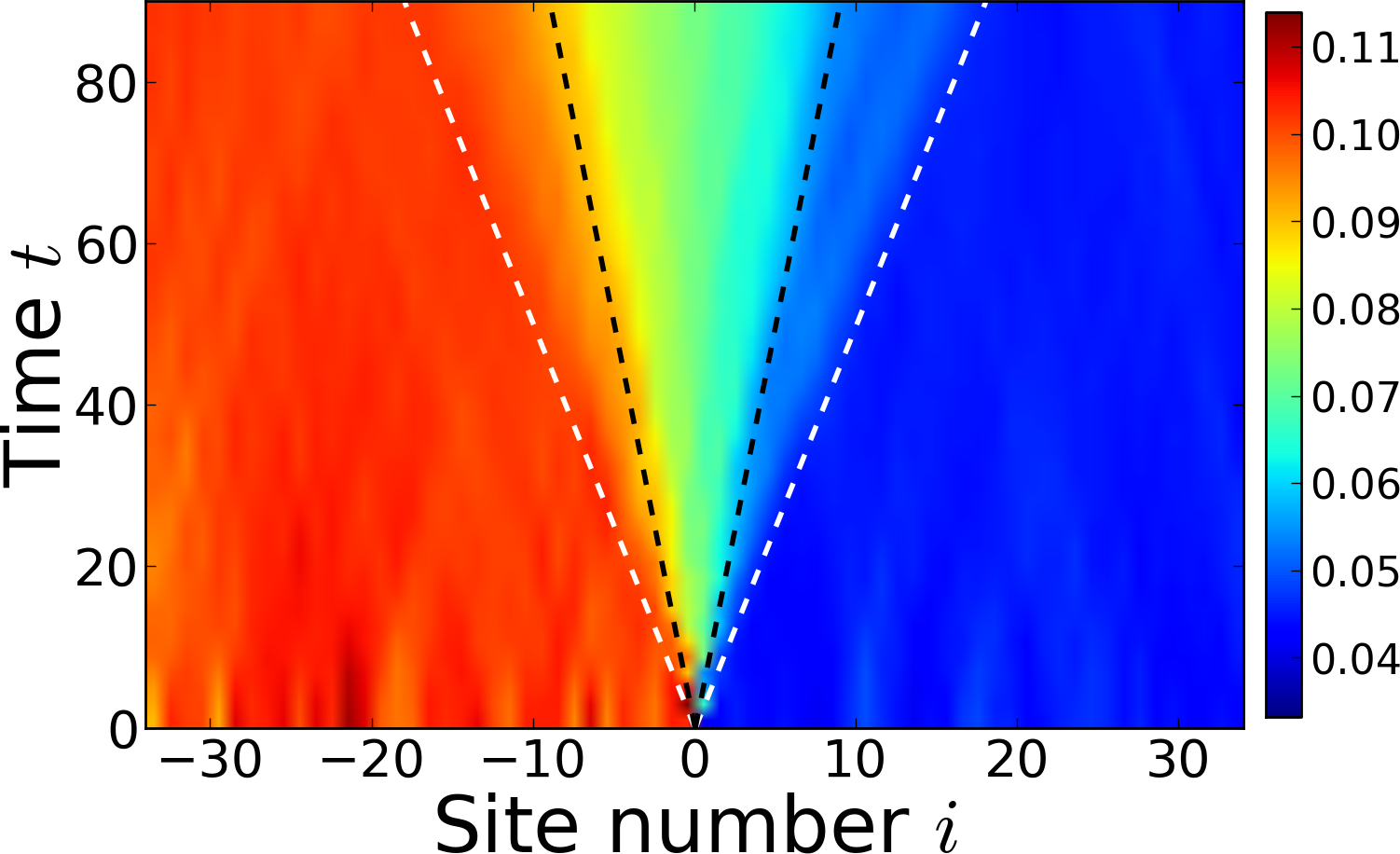}\vspace{2mm}\parbox[b][4.5cm][t]{1.6cm}{\Large$\Var{(n_i)}$}\includegraphics[width=8cm]{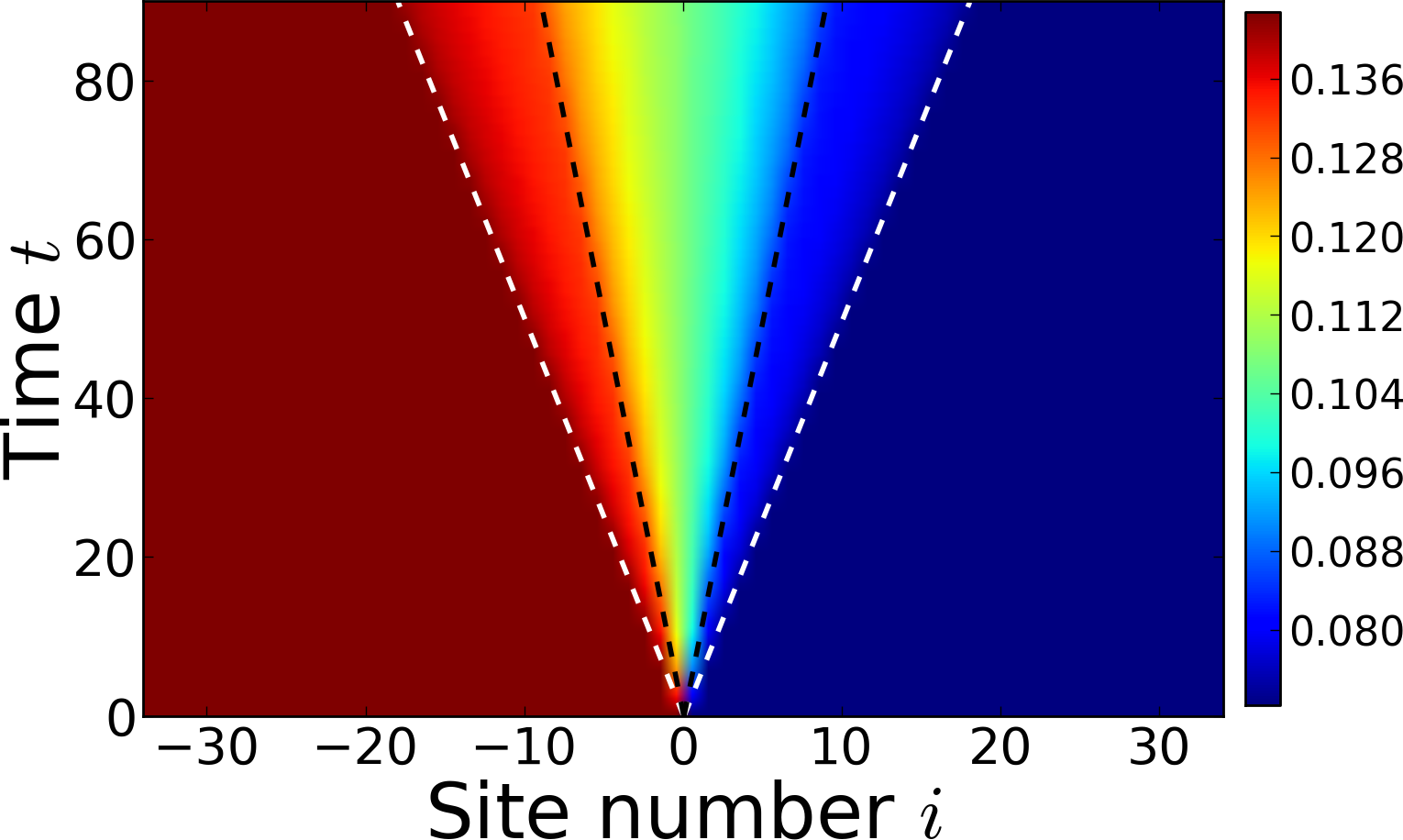} 
\includegraphics[width=8cm]{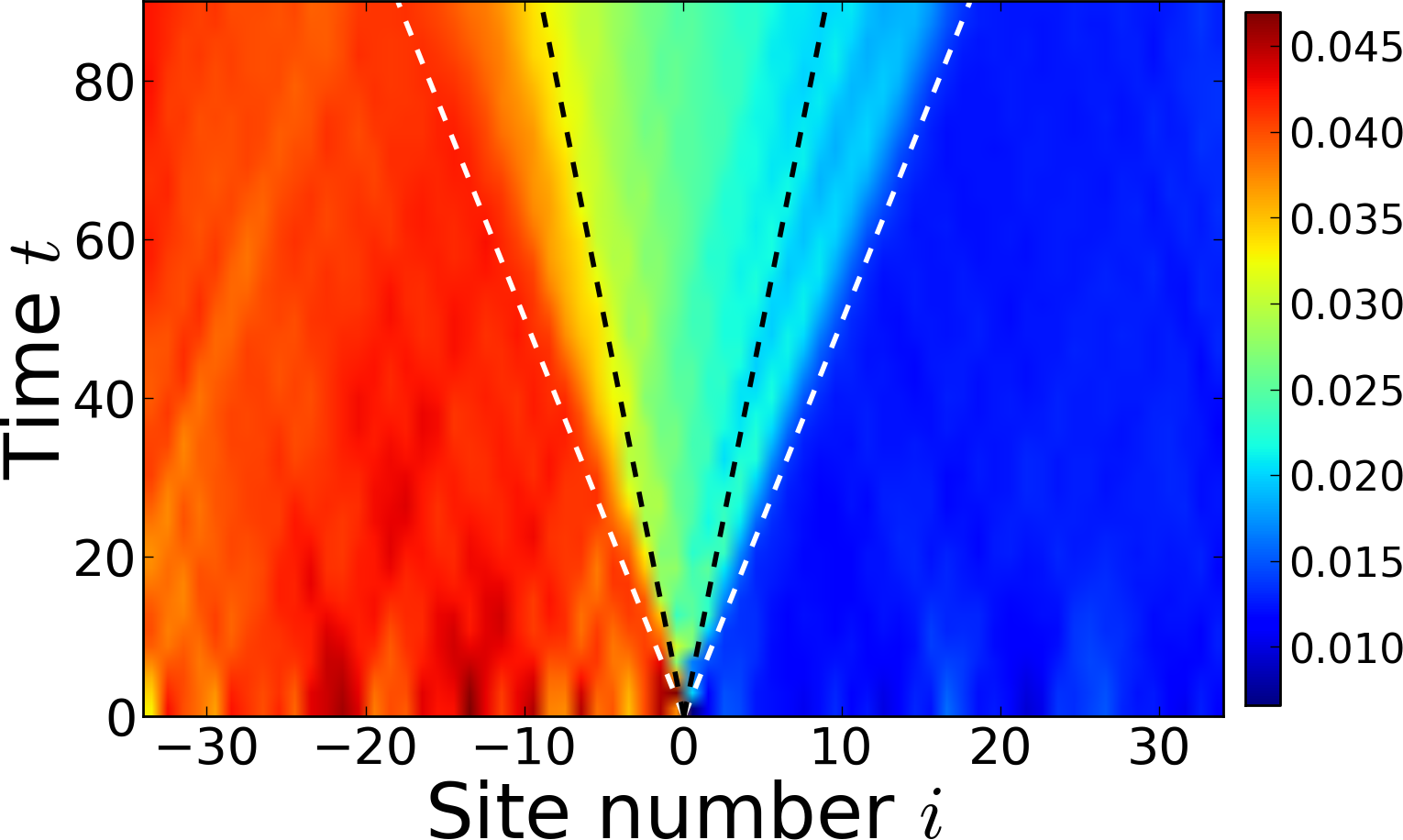}\vspace{2mm}\parbox[b][4.5cm][t]{1.6cm}{\huge$\nqp_i$}\includegraphics[width=8cm]{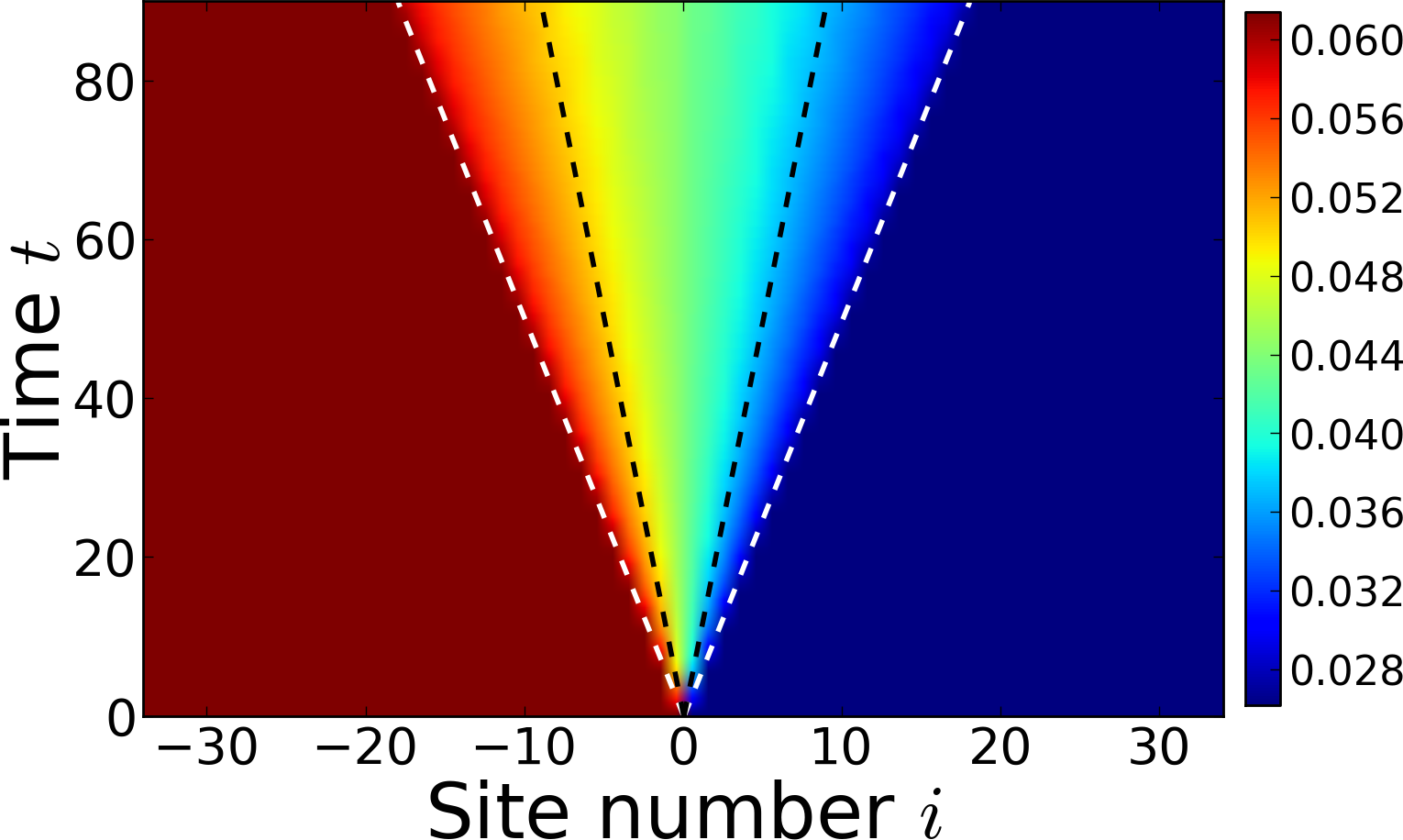} 
\includegraphics[width=8cm]{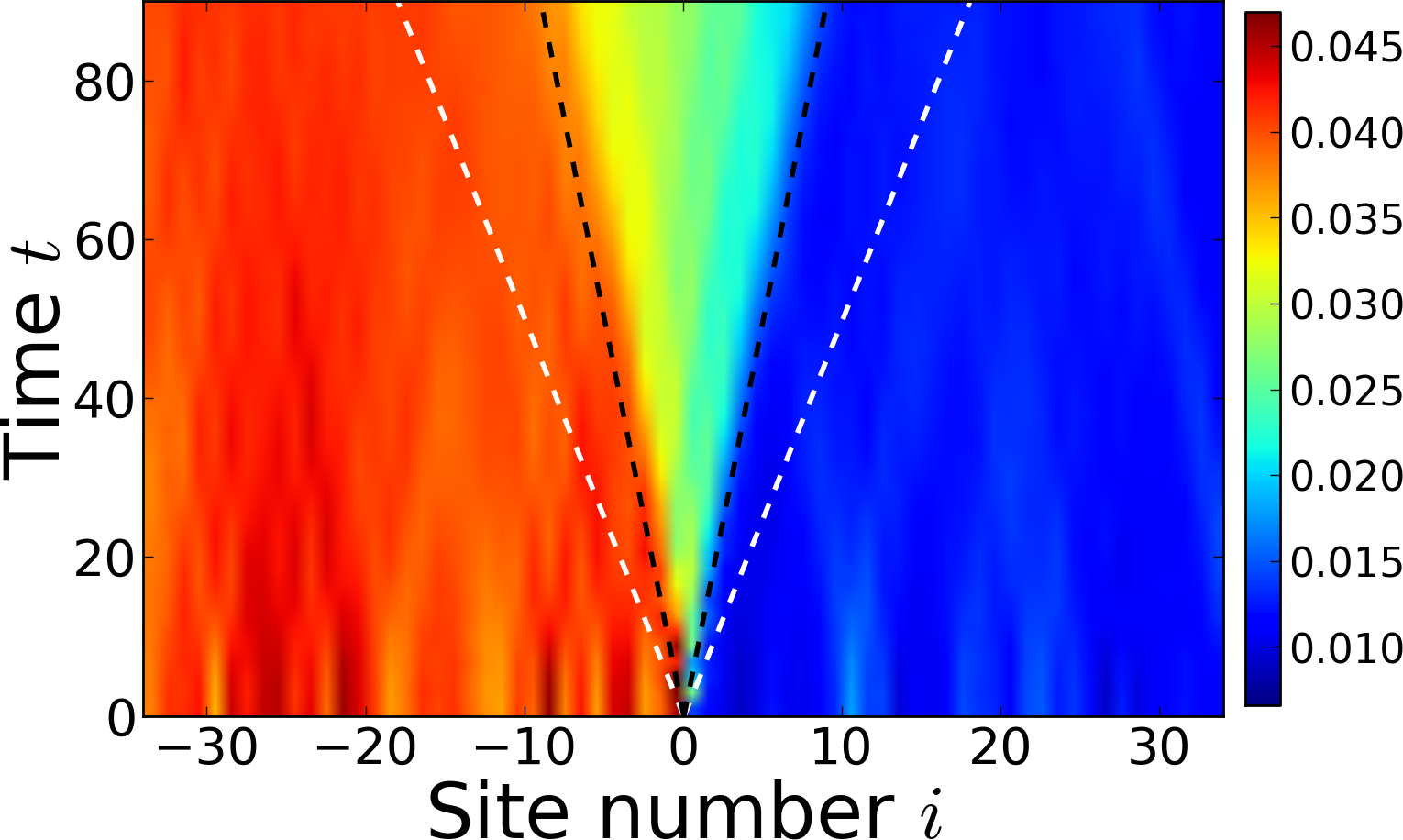}\parbox[b][4.5cm][t]{1.6cm}{\huge$\nqh_i$}\includegraphics[width=8cm]{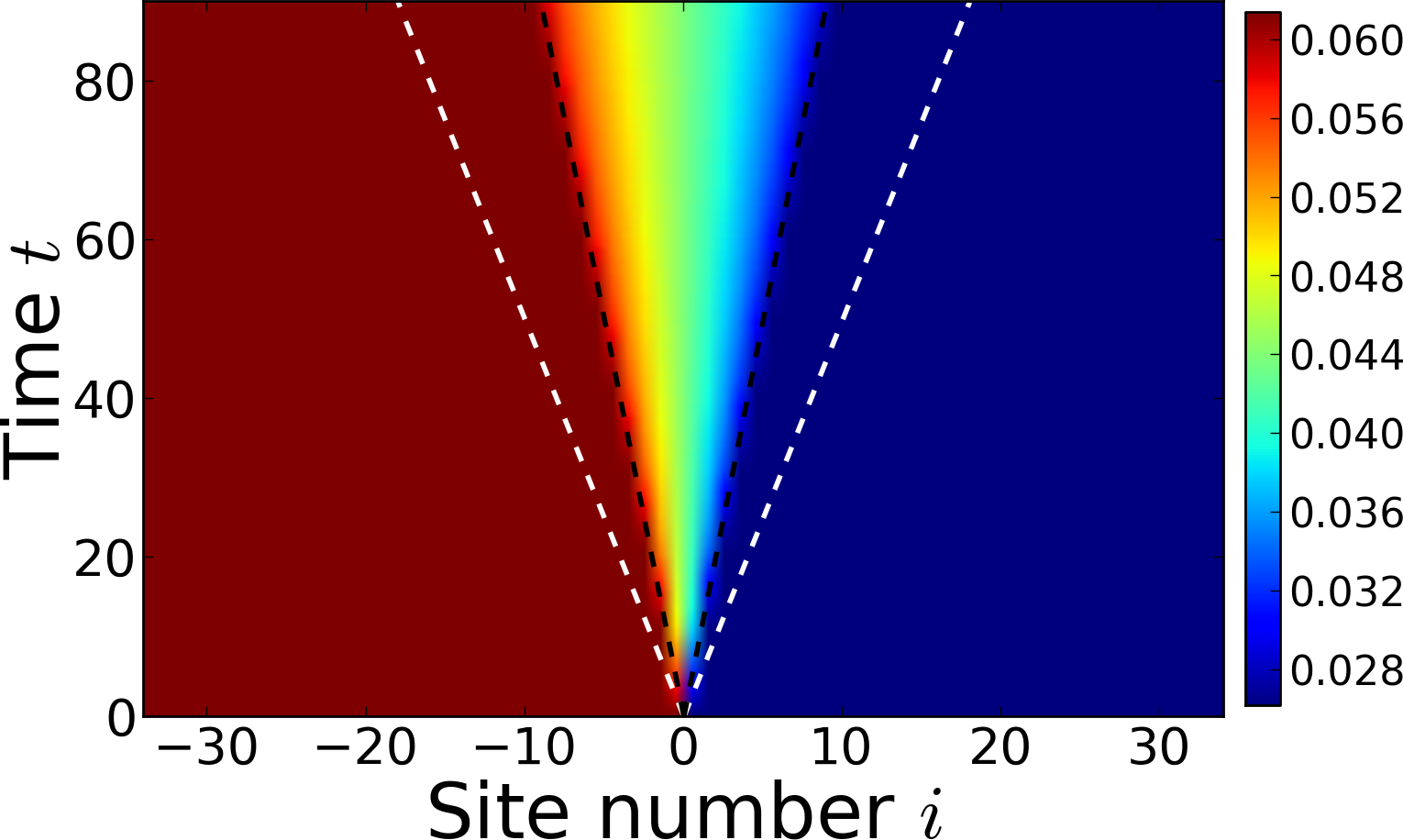} 
\caption{(color online) Comparison of {quasi-exact} numerical METTS simulations (left column) with the Bogoliubov theory (right). 
From top to bottom: mean site occupation $\avn$, variance $\avnsq-\avn^2$, quasiparticle density $\nqp_i$ and quasihole density $\nqh_i$. 
The dashed white (resp. black) straight lines correspond to ballistic propagation of quasiparticles 
(resp. quasiholes) from the connection point between the two subsamples at the maximal allowed velocity,
i.e. $x^{\mathrm{(qp)}}(t)=\pm 4Jt$ (resp. $x^{\mathrm{(qh)}}(t)=\pm 2Jt$). See text for a detailed discussion.
} 
\label{copar1}
\end{figure*}

\begin{figure*}
\includegraphics[width=8cm]{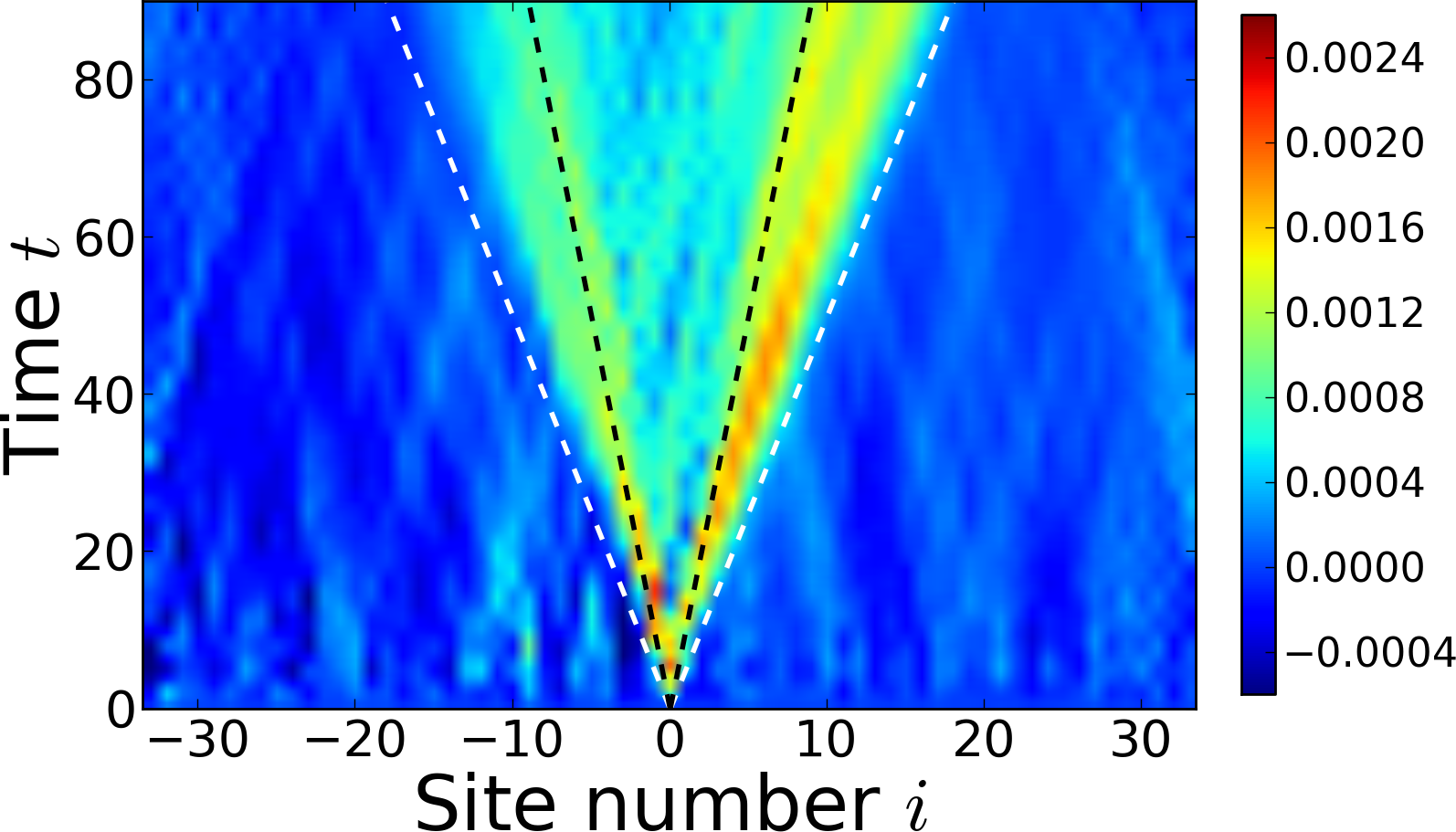}\vspace{2mm}\parbox[b][4.5cm][t]{1.6cm}{\Large$j^n(i)$}\includegraphics[width=8cm]{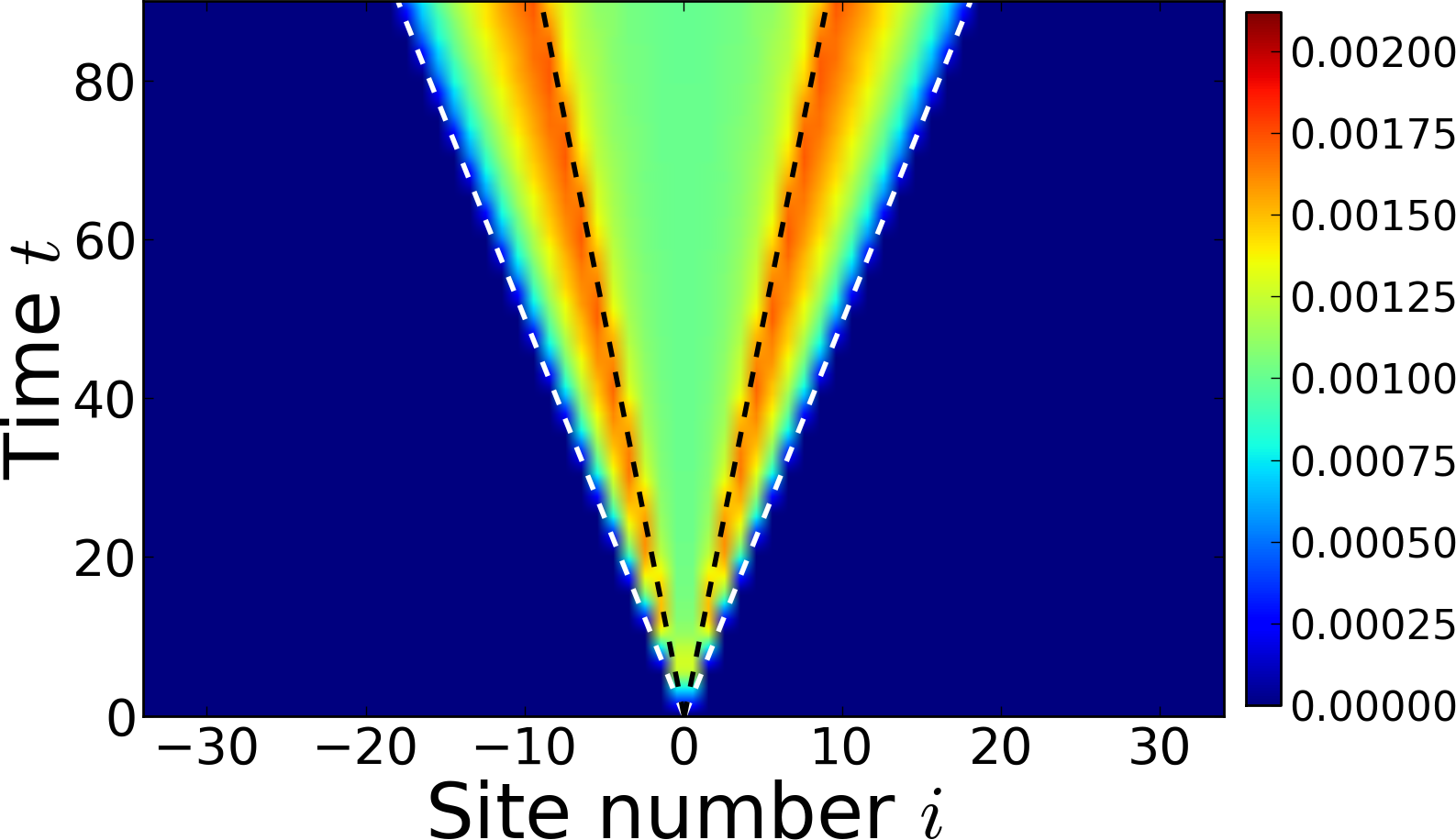} 
\vspace{1truemm}\includegraphics[width=8.cm]{current_energy_merge_68_mateusz.png}\vspace{2mm}\parbox[b][4.5cm][t]{1.6cm}{\Large$j^E(i)$}\includegraphics[width=8.cm]{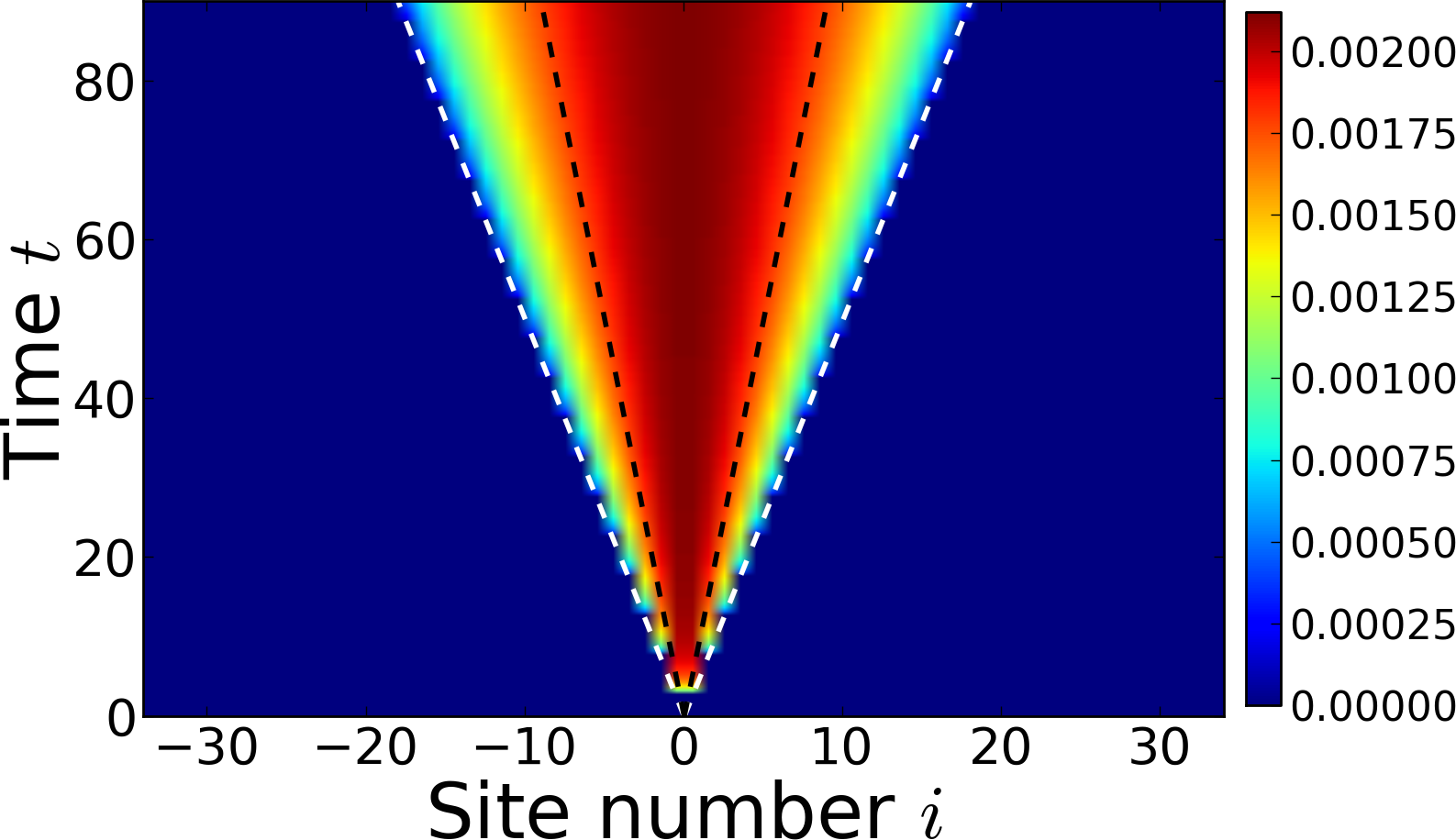} 
\vspace{1truemm}\includegraphics[width=8cm]{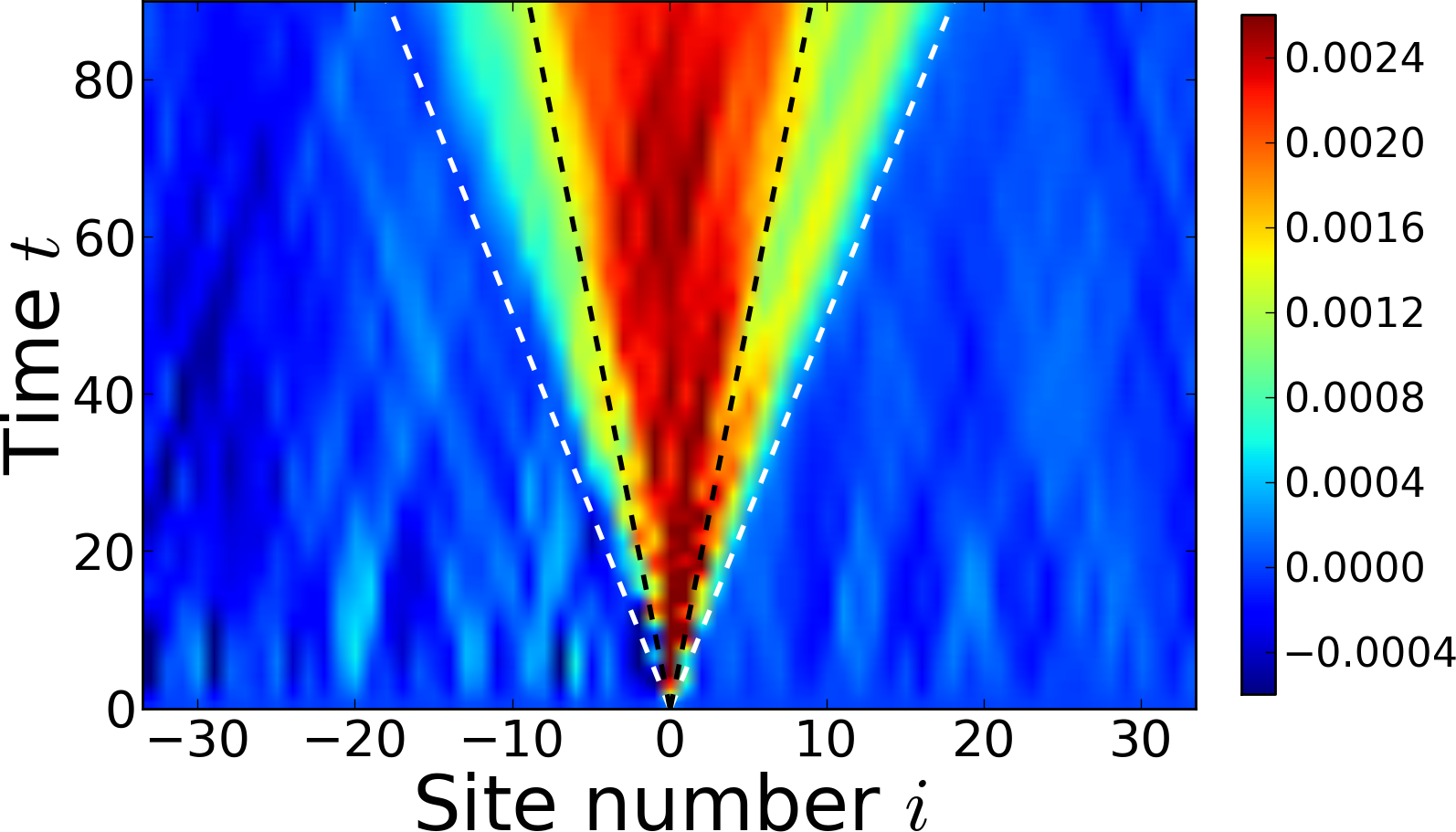}\vspace{2mm}\parbox[b][4.5cm][t]{1.6cm}{\Large$j^{\mathrm{Var}}(i)$}\includegraphics[width=8cm]{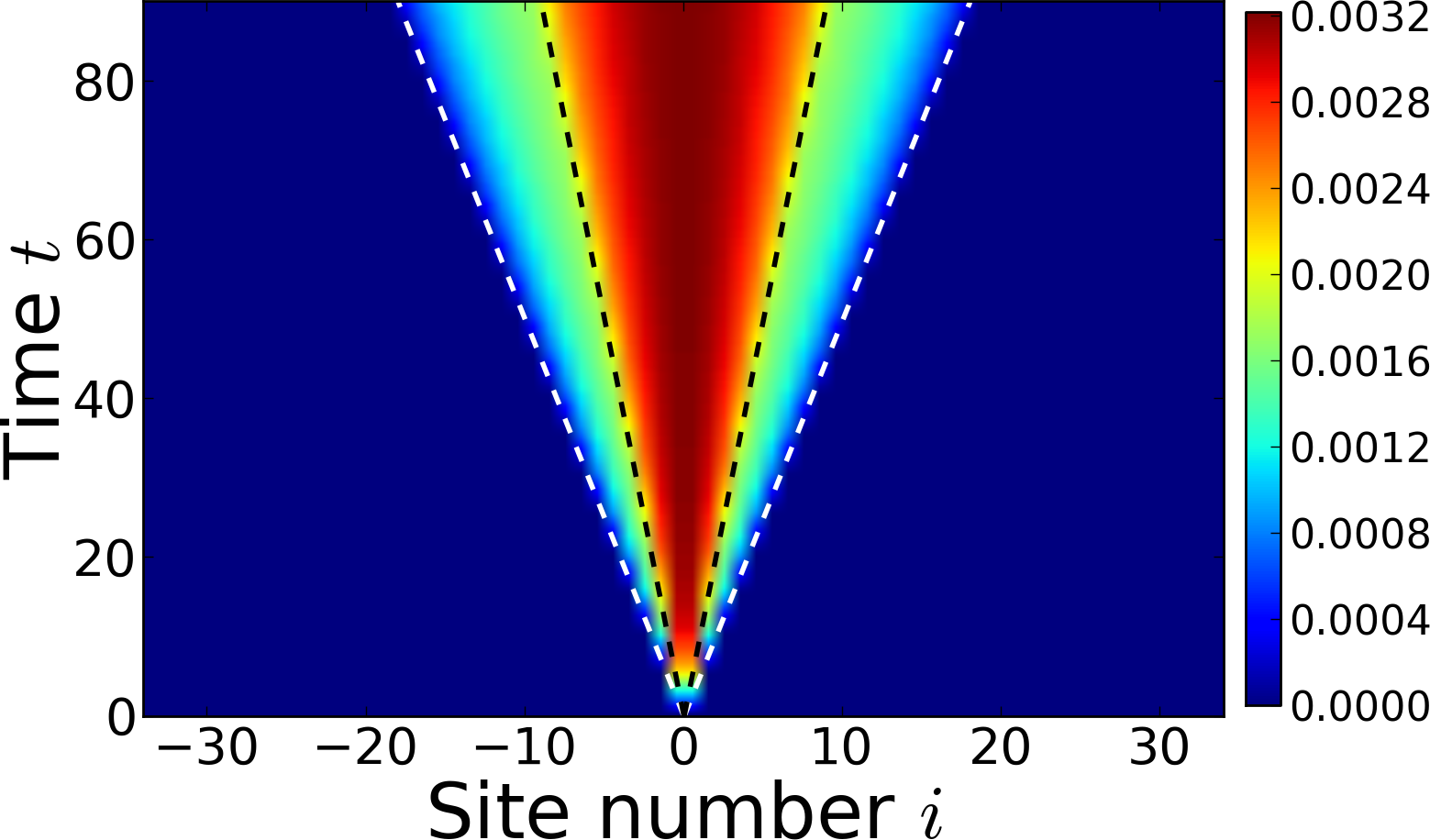} 
\vspace{1truemm}\includegraphics[width=8.cm]{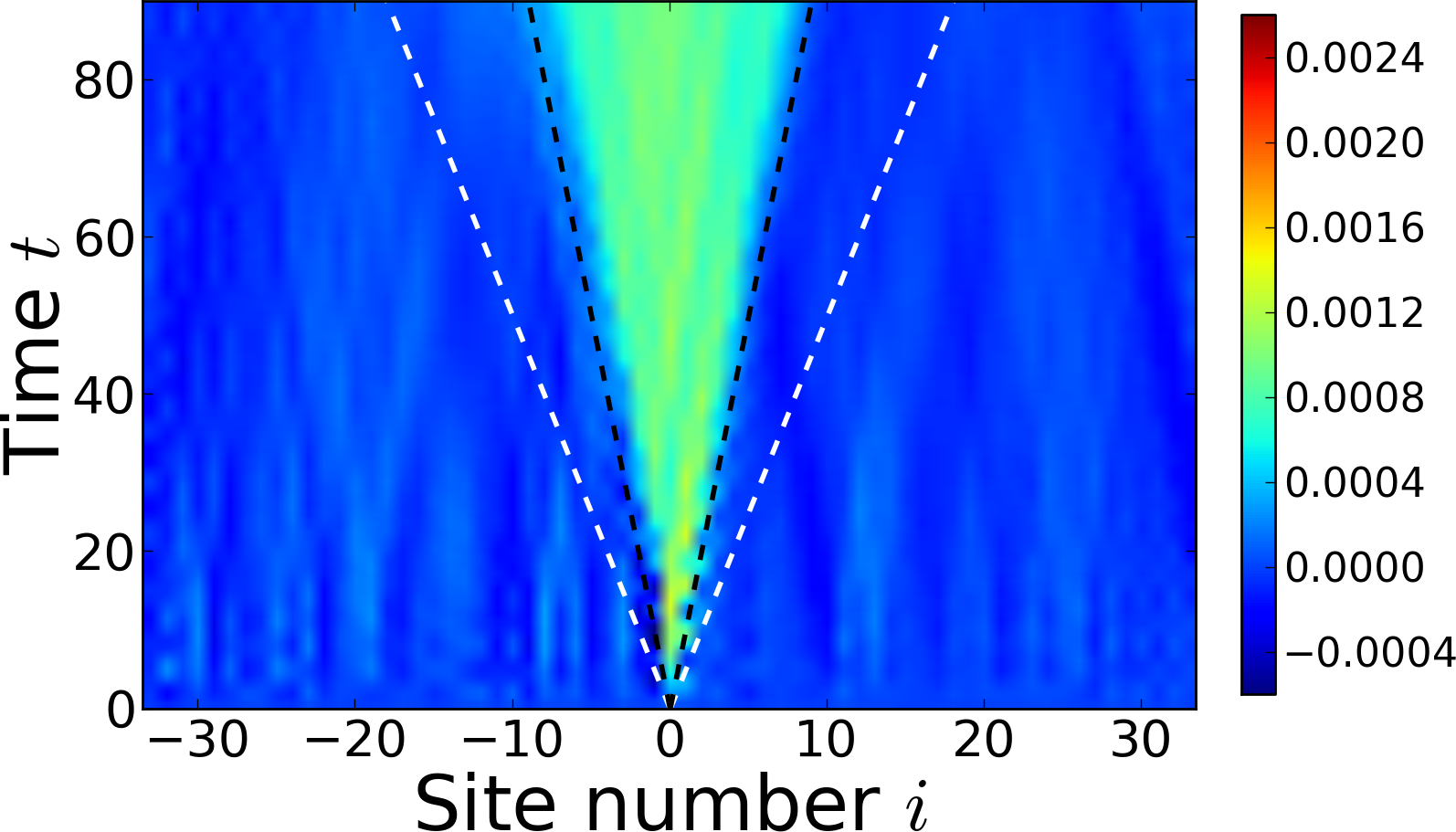}\vspace{2mm}\parbox[b][4.5cm][t]{1.6cm}{\Large$j^{(\mathrm{qh})}(i)$}\includegraphics[width=8.cm]{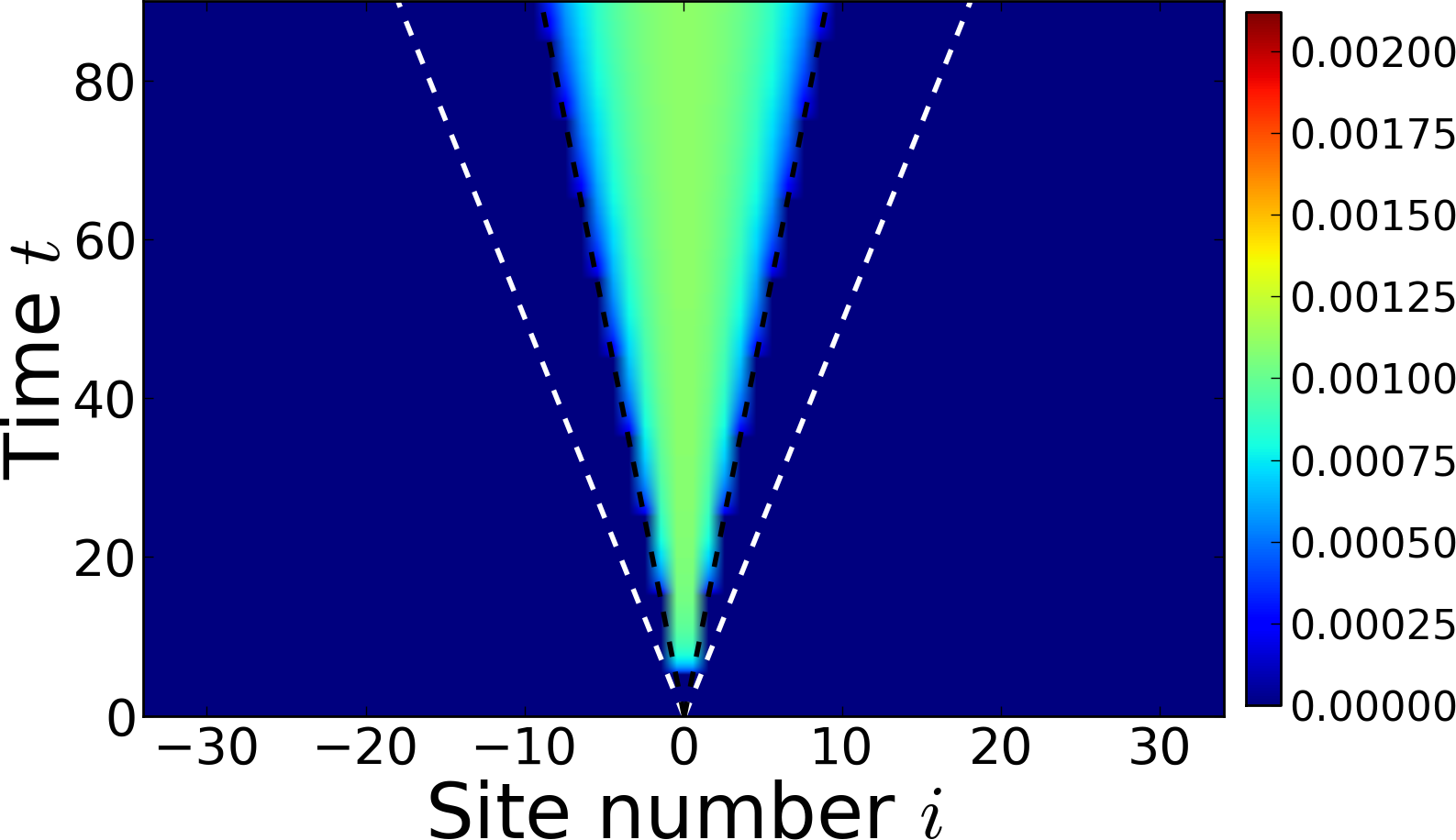} 
\caption{(color online) 
Comparison of the currents computed from the numerical METTS simulations (left column) with the Bogoliubov theory (right). 
{The currents are obtained by numerical calculations for the same parameter values as in Fig.~\ref{copar1}.} 
From top to bottom: mass current, energy current, heat (variance) current as defined in Eq.~(\ref{varcur}) and quasihole current
as defined in Eqs.~(\ref{eqcur}) (the quasiparticle current coincides with the energy current). See text for discussion.
} 
\label{coparcur}
\end{figure*}

\subsection{Sharp gluing}

We now give a more detailed analysis of the simulations already presented in Fig.~\ref{fig:fluct} and Figs.~\ref{merge1}-\ref{fig:3D1}. Recall that we consider two  Mott insulators of length $M=35,$ each one at thermal equilibrium, but with different temperatures $\beta_L U=6,\beta_R U=8$. Each insulator is represented by an appropriate thermal density matrix (METTS ensemble), the total state of the system being the tensor product of those density operators. At time $t=0$ the tunneling between touching sites is turned on to a common value $J/U=0.05$.  While in Figs.~\ref{fig:fluct} and \ref{merge1} we have shown the
average particle density $\av{n_i}$, squared particle density $\av{n_i^2}$ and variance $\Var{(n_i)}=\av{n_i^2}-\av{n_i}^2$ at selected times, Fig.~\ref{copar1} presents
the same data resulting from a quasi-exact numerical simulation as 
color plots (left column) together with the predictions given by the Bogoliubov theory (right column) described in the previous Section.

A good agreement between the quasi-exact METTS simulations and the Bogoliubov prediction is visible already in the top row, where we compare densities, $\avn$: a bump propagating to the right side, a hole propagating to the left side. On the left side, the
higher temperature implies a higher density of quasiparticles and quasiholes. Once the two parts are connected, the quasiparticles
propagate ballistically at twice the speed of the quasiholes. It results in an excess of quasiparticles -- hence a higher density -- between
the quasihole light cone (dashed black line) and the quasiparticle light cone (white dashed line).
As noted earlier, the variance $\Var{(n_i)}$, shown in the second row, is less sensitive to noise, and a very good agreement
between numerical METTS simulations and the Bogoliubov theory is observed.

While the basic ingredients of the theoretical approach are the quasiparticle and quasiholes densities, these
quantities cannot be directly measured in the METTS simulations. It is, however, possible to invert Eqs.~(\ref{Eq:occupation_prediction})
and to deduce the densities from the measured quantities $\avn$ and $\avnsq$:
\begin{eqnarray}
 \nqp_i & = & \frac{\Var{(n_i)}+\av{n_i}-1}{2} - \frac{4J^2}{U^2}\nonumber \\
 \nqh_i & = & \frac{\Var{(n_i)}-\av{n_i}+1}{2} - \frac{4J^2}{U^2}. 
\end{eqnarray}
These formulae allow us to determine the distribution in time of quasiparticles and quasiholes for simulations, as well as conversely find the predictions for the particle density and its variance from the quasiparticle distributions. 
Clearly the excess of particles is observed on the cold side (with the corresponding hole on the hot side  due to conservation of the total number of particles) moving out ballistically from the {junction} point. That is due to the spread of quasiparticles, as described in the previous section. 

Observe that while we keep the same scale for the density, we use different scales for the variance (second row in Fig.~\ref{copar1})
and for the quasiparticle (third row) and quasihole (fourth row) densities. 
That is due to finite size effects discussed extensively in the previous section. Instead of fitting the correction we have chosen to plot the raw data; the correction is ``automatically'' taken into account by the rescaling of the colorbar. This provides a convincing argument that a single correction factor valid at all times is sufficient to bring the results of METTS simulations and the  theoretical  predictions  together.  While of course some differences may be visible, the overall agreement is quite remarkable showing beyond doubt that the Bogoliubov theory  describes  very well  the results of the simulations  and thus the physics of heat and mass transfer in this system.
 Note especially the clear demonstration that quasiholes propagate more slowly than quasiparticles (by a factor 2), as the
quasihole density is affected only inside the inner light cone -- marked with the black dashed lines -- while the quasiparticle density
is affected inside the outer light cone -- marked with the white dashed lines. 
 
These conclusions are further confirmed by inspection of different possible currents. As before, the simulations give us access to
mass current $\jN$, Eq.~(\ref{jN}) as well as to the energy current $\jE$, Eq.~(\ref{jEa}). Those can be related to quasiparticle and quasihole currents via 
Eqs.~(\ref{eqcur}). One can also define the variance current, which following Eq.~(\ref{Eq:occupation_prediction}) can be defined as 
\be \label{varcur}
 j^{\mathrm{Var}}(x,t)   =  \jqp(x,t) + \jqh(x,t) = \jN(x,t) + \frac{2\jE(x,t)}{U} 
\ee
Since the variance is directly related to temperature, the variance current can  be considered as a heat current. 

Observe in Fig.~\ref{coparcur} that, as expected, the currents are nonzero only in the ``light-cone'' region around the point of junction.
The agreement between the simulations and the theory is less spectacular for mass and energy currents while it is much better for the heat current and, in particular, quasihole current (bottom row). Indeed the cone both for simulations and theory is then 
restricted by the maximal velocity of quasiholes. The differences in results between the two approaches may be again traced back to finite size effects.

\begin{figure}
\includegraphics[width=8cm]{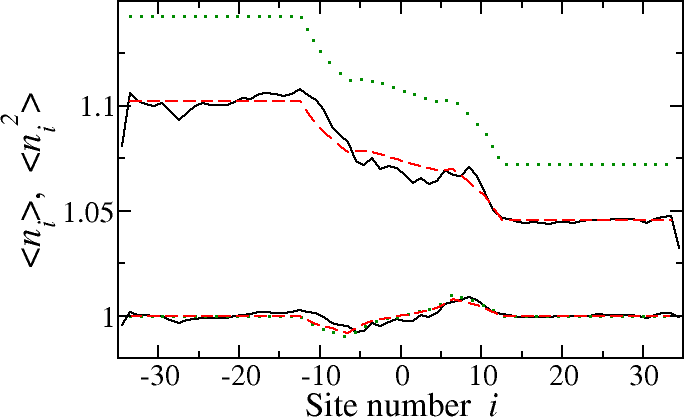}
\caption{(color online) Comparison of the average particle density $\avn$ (lower curves) and the average squared density $\avnsq$ (upper curves) at time $tU=60,$ for the quasi-exact METTS simulations (solid black lines) and the Bogoliubov theory. The raw Bogoliubov predictions are the
dotted green curves; rescaled results, taking into account finite size effects, as shown as dashed red lines and agree very well
with the METTS simulations. See text for discussion of the various features.}
\label{fig:n_n2_time_60}
\end{figure}

\begin{figure}
\includegraphics[width=8cm]{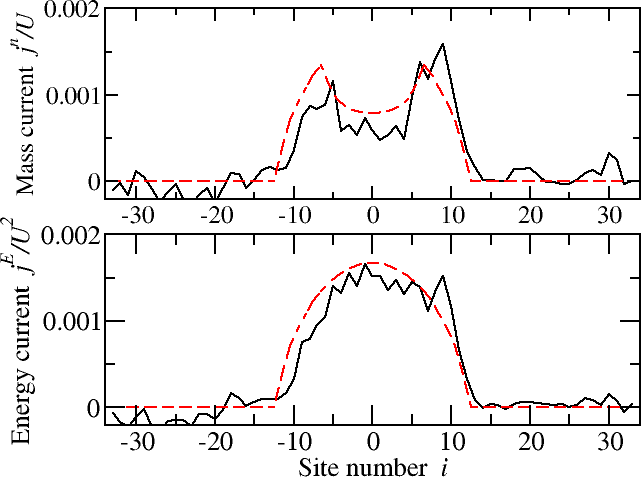}
\caption{(color online) Comparison of the mass and energy currents for the same data than in Fig.~\ref{fig:n_n2_time_60}. 
The energy current is predicted by the Bogoliubov theory to depend only on the quasiparticle current, see Eq.~(\ref{eqcur}).
It displays the characteristic ''semi-circle'' shape predicted by our Bogoliubov theory, Eq.~(\ref{eq:semi-circle}).
In contrast, the mass current is the difference between the quasiparticle and quasihole ''semi-circles'', see Eq.~(\ref{eqcur}),
and presents two maxima propagating at the maximum velocity of the quasiholes $\pm2J.$}
\label{fig_curr_t60}
\end{figure}

Color plots in Figs.~\ref{copar1},\ref{coparcur} show that our Bogoliubov approach is qualitatively correct: heat and mass transport
in the system are conveniently described by quasiparticle/quasihole excitations that propagate ballistically. 
In order to make the comparison  more
quantitative, we show in Fig.~\ref{fig:n_n2_time_60}  the average particle density $\avn$ and the average squared density $\avnsq$ at time $tU=60.$ After the proper rescaling necessary due to finite size effects (see discussion above), the agreement is very good. 
The density bump (on the right side) and hole (on the left side) is well predicted. For the average squared density $\avnsq,$ one can see three distinct regions between the ''hot'' left plateau (not yet affected) and the ''cold'' right plateau: two rather abrupt cliffs on the edges
of the two plateaus separated by an intermediate much flatter region. The two cliffs correspond to regions already reached by quasiparticles
while the intermediate flat region is affected both by quasiparticles and quasiholes. The kinks at the frontiers between regions
correspond to quasiparticles/quasiholes with maximum velocities $4J/2J,$ respectively. They are smoothed out, but still clearly visible,
in the quasi-exact METTS simulations. Note that all quasiparticles/quasiholes in the system are thermally excited, so that the effects
observed are truly due to non-equilibrium thermodynamical properties.

In Fig.~\protect{\ref{fig_curr_t60}}, we show the comparison between the mass and energy currents as computed from the METTS simulations
and the predictions of the Bogoliubov approach. Not surprisingly, the salient features are quantitatively well predicted:
the energy current, predicted by Eq.~(\ref{eqcur}) to depend only on the quasiparticle current displays a single bump
with the characteristic ''semi-circle'' shape predicted by Eq.~(\ref{eq:semi-circle}); the mass current is the difference between the quasiparticle and quasihole currents and consequently reveals two maxima propagating at the maximum velocity of  quasiholes $\pm2J.$

\subsection{Smooth gluing}

\begin{figure}
\includegraphics[width=8cm]{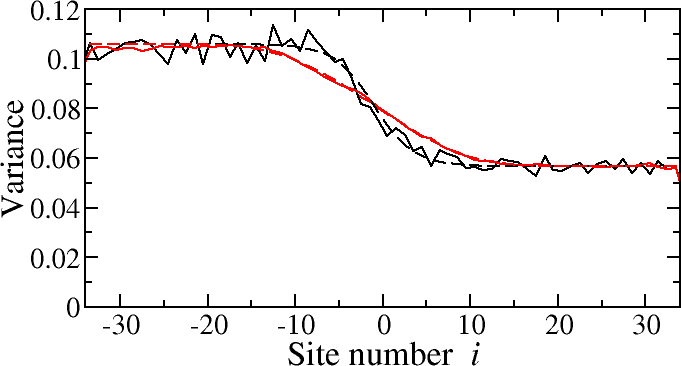}
\caption{(color online) Particle number variance at initial time $t=0$ and at $tU=60,$ for two smoothly glued Mott insulators with
(inverse) temperatures $\beta_L U=6,\beta_R U=8,$ and $J/U=0.05.$ The interface size, Eq.~(\ref{glue}), is $s=15$.
Solid lines correspond to quasi-exact numerical results using the METTS method, while dashed curves are theoretical estimates. 
A single parameter -- required to take into account finite size effects -- has been used to adjust the numerical and theoretical curves. The sharpest (black) step corresponds to the initial time $t=0.$ The smoother (red) curve corresponds to $tU=60.$ Note that the real time evolution strongly smooths the initial site-to-site fluctuations. At $tU=60,$ the theoretical prediction is almost indistinguishable from the quasi-exact numerical
simulation.}
\label{varmetts}
\end{figure}

We now confront the predictions of our Bogoliubov theory
with the numerical data obtained for a smoothly glued sample, as the one described in section~\ref{sec:smooth}.  As previously, the whole system 
consists of  70 sites  with $\beta_L U=6$ on the hot side and $\beta_R U=8$ on the cold side. 
Fig.~\ref{varmetts} shows the variance  of the particle density  across the system at initial time as well as 
after some real time evolution.  {In order} to compare the theory and the numerical experiment correctly, one has to introduce
some  rescaling, taking care of the finite size effect.   While two different rescaling factors were used for the ''sharp gluing'' case
where the two subsamples are initially not connected, we here use a \textit{single} rescaling parameter for the whole sample, and all times. 
 This is because --  by construction --  a smooth gluing procedure find the equilibrium initial state for the whole sample at the initial time. 
As before, real time evolution washes out site to site fluctuations of the variance: starting from, say $tU=10,$ 
the variance is {perfectly} reproduced by the Bogoliubov theory.

\begin{figure}
\includegraphics[width=8cm]{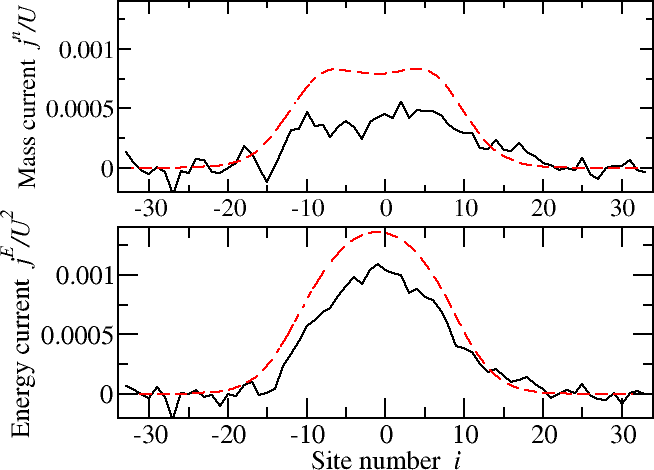}
\caption{(color online) Mass (upper panel) and energy (lower panel) currents at  $tU=60$ for the same data as in Fig.~\ref{varmetts}. 
Numerical results (solid black curves) are compared with the predictions of the Bogoliubov theory (dashed red curves). 
While the magnitude of the currents is not predicted properly, the overall shape is faithfully predicted by the theory.}
\label{curmetts}
\end{figure}

Fig.~\ref{curmetts}  presents the mass and the energy currents at  $tU=60$ using the same single scaling factor. Clearly some  quantitative discrepancies between the numerics and theoretical simulations exist. One may, however, observe quite nice qualitative shape agreement between the two curves.

\begin{figure}
\includegraphics[width=8cm]{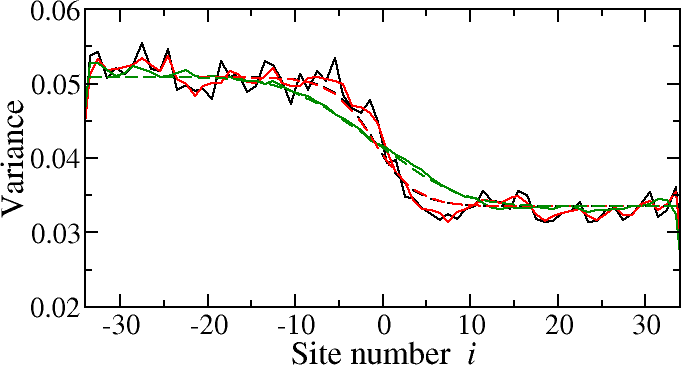}
\caption{(color online) Particle number variance at initial time $t=0$, $tU=10$ and at $tU=60,$ for two smoothly glued Mott insulators with
(inverse) temperatures $\beta_L U=8,\beta_R U=10,$ and $J/U=0.05.$ The interface size, Eq.~(\ref{glue}), is $s=15$.  Solid lines correspond to quasi-exact numerical results using the METTS method, while dashed curves are theoretical estimates. 
A single parameter -- required to take into account finite size effects -- has been used to adjust the numerical and theoretical curves. The sharpest (black) step corresponds to the initial time $t=0.$ The smoother (red and green) curves corresponds to $tU=10$ and $tU=60$. The real time evolution strongly smooths the initial site-to-site fluctuations. At $tU=60,$ the theoretical prediction is almost indistinguishable from the quasi-exact numerical
simulation.}
\label{vmetts10}
\end{figure}

\begin{figure}
\includegraphics[width=8cm]{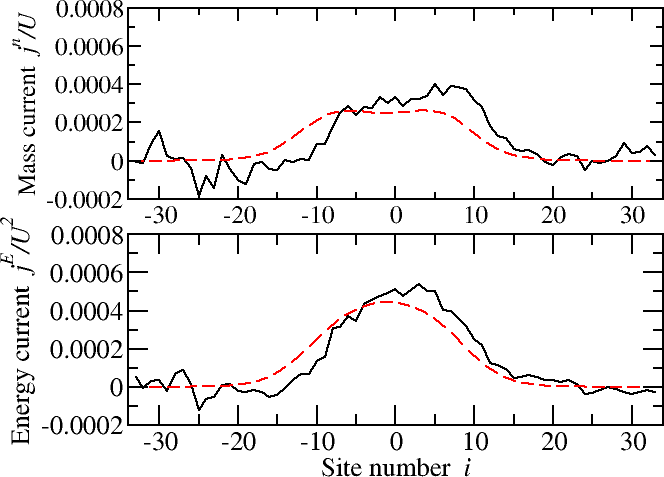}
\caption{(color online) Mass (upper panel) and energy (lower panel) currents at  $tU=60$ for the same data as in Fig.~\ref{vmetts10}. 
Numerical results (solid black curves) are compared with the predictions of the Bogoliubov theory (dashed red curves). 
The agreement is very satisfactory.  }
\label{cmetts10}
\end{figure}

As a final example, we consider the same system but at lower temperature. This time the hot part corresponds to $\beta_L U=8$ while the cold part to $\beta_R U=10,$ again with a smooth gluing of the two subsystems. 
The results are presented in Fig.~\ref{vmetts10} and Fig.~\ref{cmetts10}. While the fluctuations are notably larger (quantum effects are more visible), the behavior of the system is similar to the previously considered case. Note also that the imperfect quantitative agreement for the currents observed at higher temperature, Fig.~\ref{curmetts}, is here significantly better. A possible explanation is that, at higher temperature, our simple Bogoliubov theory breaks down at shorter
times, either because of additional excitations or because of the finite lifetime of quasiparticle/quasihole excitations.

\section{Conclusions}
\label{conclusions}
We have studied non-equilibrium dynamics occurring when two quantum insulators at different temperatures are brought into contact.
The process, in the long run, should lead to equilibration of temperatures (at least in the thermodynamic limit).
Using Mott insulators of the Bose-Hubbard model as a specific example -- as it is close to experimental realization with ultracold atoms --
we, however, observe that heat transport is initially ballistic: it efficiently transfers energy from the hot to the cold side of the system, but the full thermalization may occur only on a longer time scale, not reachable by our numerical simulations.

Fully numerical, ``quasi-exact'' calculations have been performed using the METTS algorithm~\cite{White09}.
Even with important computer resources, we could only study the dynamics on a rather limited amount of time.
This is due to two primary reasons. First, the METTS ensemble needed to obtain reasonable averages consisted of more than 10 000 vectors; each of them had to be evolved using the TEBD (or t-DMRG)  algorithm. The price to pay is a direct $10^4$ increase of the computational time in comparison to $T=0$ studies of e.g. quantum quenches. Second, for $T>0,$ the METTS ensemble includes by construction some significantly excited states; the real time evolution of such states leads to a fast increase of the entanglement, requiring significant extension of the bond dimension in the MPS description. \textit{A posteriori,} one may conclude that this is not the method of choice and other approaches such as a purification scheme \cite{Karrasch13} or methods employing Heisenberg 
picture evolution \cite{Pizorn14} could perform much better, at least according to comments in these papers. 
A very recent comparison of purification schemes with METTS  \cite{Barthel14} reached similar conclusions. 

The numerical results have been quantitatively compared with a simple analytic  theory based on Bogoliubov excitations. 
The theory explains the ballistic transport numerically observed in our system providing good estimates for observable quantities, provided some plausible corrections are included that took into account the finite size effects.   
The theory applies also for smoothly glued system: the transport may be fully understood as the movement of quasiparticles and quasiholes of the integrable Bogoliubov theory.  
In some sense, this explains the success of the METTS approach for description of this particular system. As conjectured by Prosen \cite{Prosen07}, the entanglement growth in an integrable system is slower and easier to handle.

\acknowledgments 
The authors wish to thank J. Dziarmaga for enlightenment on quasiparticles. We acknowledge the use of the ALPS library~\cite{Bauer:Alps2:JSM11} for Quantum Monte-Carlo calculations using the worm 
and directed loop algorithms.
This work has been supported by Polish National Science Center within project No. DEC-2012/04/A/ST2/00088.  M.\L. acknowledges support of the Polish National Science Center by means of project no. 2013/08/T/ST2/00112 for the PhD thesis. Numerical simulations were performed thanks to the PL-Grid project: contract number: POIG.02.03.00-00-007/08-00 and at Deszno supercomputer (IF UJ) obtained in the framework of the Polish Innovation Economy Operational Program (POIG.02.01.00-12-023/08).
This work was performed within Polish-French bilateral program POLONIUM No. 33162XA.

%\bibliography{finiteT}


\end{document}